\documentclass[compsoc]{IEEEtran}
\usepackage{amsmath,amsfonts}
\usepackage{algorithmic}
\usepackage{array}
\usepackage[caption=false,font=normalsize,labelfont=sf,textfont=sf]{subfig}
\usepackage{textcomp}
\usepackage{stfloats}
\usepackage{url}
\usepackage{verbatim}
\usepackage{graphicx}
\def\BibTeX{{\rm B\kern-.05em{\sc i\kern-.025em b}\kern-.08em
    T\kern-.1667em\lower.7ex\hbox{E}\kern-.125emX}}
\usepackage{balance}
\usepackage{xspace} 
\usepackage{hyperref}
\usepackage{enumitem}
\usepackage{changepage}

\newcommand{\finding}[2]{
\begin{adjustwidth}{3mm}{3mm}
\vspace{2.5mm}
\noindent
\textbf{\textit{Finding #1. }} \textit{
#2
}
\vspace{2.5mm}
\end{adjustwidth}
}

\def\brush{Distortion-aware brushing\xspace}

\usepackage{mathptmx}                  

\usepackage{color}
\usepackage{xcolor}
\usepackage{amsmath}
\DeclareMathOperator*{\argmax}{arg\,max}

\usepackage{setspace}
\setlist{leftmargin=3.5mm}

\newcommand{\close}[2]{\texttt{close}$_{\kappa}$\(({#1}, {#2})\)}
\newcommand{\simil}[2]{\texttt{sim}$_k$\(({#1}, {#2})\)}
\newcommand{\dens}[1]{\texttt{dens}\(({#1})\)}

\newcommand{\boldsubsubsection}[1]{
\vspace{4pt}
\noindent
\textbf{#1.}
}

\definecolor{applepinknormal}{RGB}{255, 55, 95}

\newcommand{\revise}[1]{#1}
\newcommand{\reviset}[1]{#1}

\definecolor{mygreen}{RGB}{102,194,165}
\definecolor{myorange}{RGB}{252,141,98}
\definecolor{myblue}{RGB}{141,160,203}
\definecolor{mypink}{RGB}{231,138,195}
\definecolor{mylgreen}{RGB}{166,216,84}
\definecolor{blueorm}{RGB}{65,129,195}

\usepackage{longfbox} 

\makeatletter
\newdimen\@tempdimd
\makeatother

\newfboxstyle{patternparam}{padding-top=2pt, padding-bottom=3pt, padding-left=3pt, padding-right=2pt, border-style=none, height=6.5pt, border-radius=3pt}

\newfboxstyle{patternparamlong}{padding-top=2pt, padding-bottom=3pt, padding-left=3pt, padding-right=3.3pt, border-style=none, height=6.5pt, border-radius=3pt}

\newfboxstyle{patternparamsolid}{padding-top=2pt, padding-bottom=4.3pt, padding-left=3pt, padding-right=3pt, border-style=none, height=6pt, border-radius=3pt}

\newcommand{\oone}{\lfbox[patternparam, background-color=myblue!90]{{\color{white}{{\normalfont \textsf{O1}}}}}}

\newcommand{\otwo}{\lfbox[patternparamlong, background-color=myblue!90]{{\color{white}{{\normalfont \textsf{O2}}}}}}

\newcommand{\othree}{\lfbox[patternparamlong, background-color=myblue!90]{{\color{white}{{\normalfont \textsf{O3}}}}}}

\newcommand{\ofour}{\lfbox[patternparamlong, background-color=myblue!90]{{\color{white}{{\normalfont \textsf{O4}}}}}}

\newcommand{\stepone}{\lfbox[patternparamsolid, background-color=myorange!40]{{\color{black}{{\normalfont \textsf{Step 1}}}}}}

\newcommand{\steptwo}{\lfbox[patternparamsolid, background-color=myorange!40]{{\color{black}{{\normalfont \textsf{Step 2}}}}}}

\newcommand{\stepthree}{\lfbox[patternparamsolid, background-color=myorange!40]{{\color{black}{{\normalfont \textsf{Step 3}}}}}}

\newcommand{\stepfour}{\lfbox[patternparamsolid, background-color=myorange!40]{{\color{black}{{\normalfont \textsf{Step 4}}}}}}

\begin{document}

\title{Distortion-aware Brushing for Reliable Cluster Analysis in Multidimensional Projections}
\author{ Hyeon Jeon, Micha\"el Aupetit, Soohyun Lee, Kwon Ko,\\ Youngtaek Kim, Ghulam Jilani Quadri, and Jinwook Seo
\IEEEcompsocitemizethanks{
    \IEEEcompsocthanksitem Hyeon Jeon, Soohyun Lee, and Jinwook Seo are with Seoul National University. 
    E-mail: \{hj, shlee\}@hcil.snu.ac.kr, jseo@snu.ac.kr
    \IEEEcompsocthanksitem Micha\"el Aupetit is with Qatar Computing Research Institute, Hamad Bin Khalifa University. 
    E-mail: maupetit@hbku.edu.qa
    \IEEEcompsocthanksitem Kwon Ko is with Stanford University.
    E-mail: hyungkwonko@gmail.com
    \IEEEcompsocthanksitem Youngtaek Kim is with Samsung Electronics. \\ 
    E-mail: {ytaek.kim@hcil.snu.ac.kr} 
    \IEEEcompsocthanksitem Ghulam Jilani Quadri is with the University of Oklahoma. \\
    E-mail: {quadri@ou.edu}
    \IEEEcompsocthanksitem Jinwook Seo and Micha\"el Aupetit are corresponding authors.
    \protect\\
}}

\markboth{IEEE Transactions on Visualization and Computer Graphics}%
{}

\IEEEtitleabstractindextext{
\begin{abstract}
Brushing is a common interaction technique in 2D scatterplots, allowing users to select clustered points within a continuous, enclosed region for further analysis or filtering. 
However, applying conventional brushing to 2D representations of multidimensional (MD) data, i.e., Multidimensional Projections (MDPs), can lead to unreliable cluster analysis due to MDP-induced distortions that inaccurately represent the cluster structure of the original MD data.
To alleviate this problem, we introduce a novel brushing technique for MDPs called \textit{\brush}.
As users perform brushing, \brush correct distortions around the currently brushed points by dynamically relocating points in the projection, pulling data points close to the brushed points in MD space while pushing distant ones apart. This dynamic adjustment helps users brush MD clusters more accurately, leading to more reliable cluster analysis. 
Our user studies with 24 participants show that \brush significantly outperforms previous brushing techniques for MDPs in accurately separating clusters in the MD space and remains robust against distortions.
We further demonstrate the effectiveness of our technique through \revise{two use cases}: \revise{(1) conducting cluster analysis of geospatial data and (2) interactively labeling MD clusters.}

\end{abstract}
\begin{IEEEkeywords}
Multidimensional Projections, Distortion-aware Brushing, Brushing, Distortions, Visual Clustering, Cluster Analysis
\end{IEEEkeywords}
}

\maketitle
\IEEEdisplaynontitleabstractindextext

\section{Introduction}

\label{sec:intro}
\IEEEPARstart{B}{rushing} is a process of selecting data points within a continuous region in the 2D space via direct manipulation such as dragging, clicking, or lassoing \cite{heer08chi}. This interaction technique allows users to focus on the selected points by labeling or highlighting them \cite{becker87technometrics, becker87statistical}. 
Since its initial introduction~\cite{fisherkeller75pacific}, brushing has become a common interaction method in visual analytics.
One important use of brushing is the identification and analysis of \textit{clusters} within multidimensional (MD) data through multidimensional 2D projections (MDPs)
~\cite{xia22tvcg, aupetit14vast, cavallo19tvcg, wenskovitch18tvcg, quadri21tvcg, jeon24tvcg, nonato19tvcg}.
MDPs are often created through dimensionality reduction algorithms, e.g., $t$-SNE~\cite{maaten08jmlr}, or by mapping two attributes onto the $x$ and $y$ axes (i.e., orthogonal projections).

\begin{figure*}
    \centering
    \includegraphics[width=\textwidth]{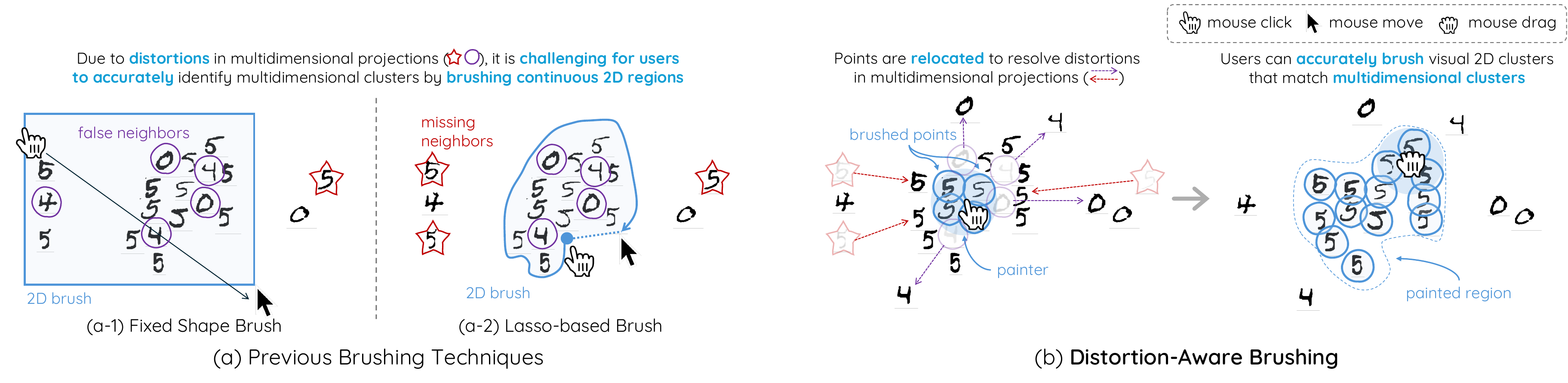}
    \vspace{-8mm}
    \caption{Comparison between existing brushing techniques (\autoref{sec:mdpbrushrel}) and \brush in identifying clusters within multidimensional (MD) data through its 2D projection. (a) Previous brushing techniques work by defining a continuous 2D region via direct manipulation (e.g., lassoing). As projections may not accurately reflect original MD data distribution due to distortions, users cannot precisely identify MD clusters.. (b) \brush supports users to precisely extract MD clusters by resolving distortions based on point relocation.}
    \label{fig:teaser}
\end{figure*}

However, visual analytics of MD data by applying conventional brushing techniques on MDPs can easily be \textit{unreliable}, i.e., insights gained from the analyses may not accurately reflect the underlying data. 
Conventional 2D brushing methods typically struggle to detect clusters in the original MD space because MDPs distort the original data \cite{aupetit07neurocomputing, lespinats11cgf, lespinats07tnn, nonato19tvcg} (\autoref{fig:teaser}a). For instance, data close in the MD space can be split apart in the 2D layout, forming missing neighbors (MN), while nearby points in the layout can come from remote regions in the MD data space, generating false neighbors (FN).  These distortions can stem from various factors, such as complex structure and high dimensionality \cite{espadoto21tvcg}, inappropriate hyperparameter selection, and inappropriate design of DR technique \cite{lee11pcs,jeon22vis} or quality metrics \cite{jeon24tvcg2}.
As a result, conventional brushing techniques might capture 2D clusters that are less cohesive or incomplete when mapped back to their original MD context.

To address this issue, several MDP brushing techniques \cite{novotny06wscg, ward94vis, martin95vis, aupetit14vast} have been proposed, yet they still face challenges with MDP distortions.
These techniques generally work by first brushing a specific 2D region and automatically mapping this selection to an MD region.
This workflow makes the final MD brushing vulnerable to distortions as it depends on a continuous 2D region that is subject to these distortions (\autoref{fig:teaser}a). 
These techniques may further constrain data analysis by using fixed shapes for the brushed regions, such as circles and hyperspheres \cite{aupetit14vast}, or rectangles and hypercubes \cite{ward94vis, martin95vis}, which cannot effectively capture clusters with non-trivial shapes in real-world datasets.

We propose \textit{\brush}, a novel brushing technique designed to overcome these issues, enabling users to more accurately identify MD clusters from their 2D projections compared to existing techniques.
Our approach addresses distortions in MDPs by persistently drawing points close in the MD space towards the brushed points and repelling those that are farther apart (\autoref{fig:teaser}b).
This relocation ensures that 2D brushes accurately mirror the composition of MD clusters in 2D space.
Therefore, our technique ensures a more reliable cluster analysis of MD data even in cases where MDPs suffer from severe distortions.

Through quantitative studies with 24 participants, we
demonstrate that \brush can accurately brush MD clusters despite distortions, surpassing previous brushing techniques for MDPs.
We also showcase how \brush can be leveraged to support \revise{cluster analysis and the interactive labeling of noisy datasets.}
We conclude the paper by discussing the benefits and limitations of \brush.

\section{Background and Related Work}
\label{sec:rel}

Our work is relevant to three areas: MDP distortions, interactive point relocation, and brushing techniques for MDPs.

\subsection{Distortions in Multidimensional Projections}

\label{sec:distortions}

MDPs aim to represent MD data in 2D space while preserving the original characteristics of the given data. 
For example, dimensionality reduction techniques are used to generate MDPs, providing a visual density-based summary of the data distribution and patterns \cite{nonato19tvcg}. 
However, MDP distortions \cite{aupetit07neurocomputing} can interfere with users' ability to analyze clusters or detect outliers in MD data  \cite{etemadpour15tvcg, brehmer14beliv}, resulting in unreliable visual analytics \cite{jeon25chi, jeon25arxiv}.

MN (Missing Neighbors) and FN (False Neighbors) \cite{ lespinats11cgf, lespinats07tnn} are typical MDP distortions that affect the 2D representation of MD data patterns \cite{nonato19tvcg}  (\autoref{fig:teaser}a). Quantitative metrics such as Trustworthiness and Continuity (T\&C) \cite{venna06nn} are commonly used in practice \cite{jeon21tvcg, lee09neurocomputing} to measure and visualize the amount of MN and FN distortions. 
MDP can be enriched to visualize the amount and type of the distortions \cite{nonato19tvcg} by coloring points \cite{martins14cg} or a region around them using heatmaps \cite{martins14cg}, Voronoi cells \cite{lespinats11cgf, heulot12infovis}, or using a network overlay \cite{martins14cg}. 
Although these approaches help explore MDP distortions and reliably analyze the structure of MD data, they are only visual indicators that can inform the brushing process but do not feature brushing actions.

\subsection{Interactive Points Relocation}

Interactive point relocation is widely adopted to explore the underlying structure of MD data. 
Dust-and-Magnet \cite{yi05dustmagnet} and iPCA \cite{jeong09ipca} allow interactive steering of the MDP layout based on the attribute values.
Another approach is to visualize MD data as snippet images and allow users to arrange MDPs by visual similarity between the snippets \cite{joia11lamp, xia23tvcg}. Yet another technique \cite{KruigerHSTH17} proposes brushing a cluster of points detected within one MDP layout, freezing it in position, and then visualizing the remaining data in the same layout using another MDP technique. 
However, these approaches focus on finding interesting visual cluster patterns under the MDP constraints rather than preventing distortions; thus, they can be used to identify interesting insights from MD data but cannot guarantee reliable cluster analysis.

Meanwhile, some previous works aimed to resolve errors locally through relocation. For example, Probing Projections \cite{stahnke16tvcg} transiently relocates points based on their MD similarity with a user-selected point so that it removes entire MN and FN distortions, but only for the selected point. 
Proxilens \cite{heulot13vamp} focuses on true MD neighbors of the selected point by pushing FN to the border of a 2D magic lens centered on that point while highlighting MN with proximity coloring \cite{aupetit07neurocomputing}. 
In contrast, instead of transiently resolving distortions around a single point, our technique maintains the correction of distortions related to a set of points, generating a persistent visual pattern (\autoref{fig:teaser}b) that enables the accurate identification and analysis of MD clusters.

\subsection{Brushing Multidimensional \revise{Data}}

\label{sec:mdpbrushrel}

We review brushing techniques for \revise{MD data} and categorize them into two groups: \textit{axis-guided brushing} and \textit{data-guided brushing}.
The former works on scatterplot matrices (SPLOMs), aiming to explore how brushed points in one orthogonal projection are distributed across other attribute spaces.
The latter works on a single MDP, where the system automatically infers the corresponding MD region based on the 2D brushed region.


\subsubsection{Axis-Guided Brushing \revise{in SPLOMs}}

\label{sec:axisguided}

The early works on brushing \revise{MD data} are designed to brush along the axes. In PRIM-9 \cite{fisherkeller75pacific}, brushing is done by adjusting the range of the 2D rectangular brush region along two axes of an orthogonal projection. Becker et al. \cite{becker87technometrics, becker87statistical} applied the same strategy to SPLOMs consisting of multiple linked orthogonal projections.
However, these techniques allow for the brushing of at most two axes, making it challenging to explore MD structures that span more dimensions.
To resolve this problem, Ward proposed $N$-dimensional brushing as a feature of XmdvTool \cite{ward94vis}, allowing users to define multiple 2D brushes within different projections of a SPLOM. 
A later version of XmdvTool \cite{martin95vis} enables users to apply logical operators (e.g., \texttt{AND}, \texttt{XOR}) between multiple MD brushed regions for more flexibility.

As these SPLOM-based techniques \revise{are designed for exploring MD data patterns across different data attribute pairs, they are ineffective for identifying MD clusters with non-linear relationships between data attributes (see Appendix  I). Moreover,} these techniques inherently rely on multiple 2D orthogonal projections, each subject to FN distortions.
Furthermore, SPLOM uses $O(M^2)$ scatterplots for representing $M$-dimensional data, making brushing hardly practical when $M$ is large. 
Parallel coordinate plots (PCP) \cite{inselberg85vc} provide an alternative axis-based representation of MD data denser than SPLOM, for which advanced techniques for brushing have been proposed \cite{roberts19pcpbrush, hauser02infovis}. However, PCP shatters the MD clusters into $M$-linked 1D projections (axes),  each generating more FN distortions than 2D projections. 

\subsubsection{Data-Guided Brushing \revise{in MDPs}}

\label{sec:dgb}

Data-guided approaches are proposed \revise{to identify MD clusters from MDPs}. These techniques follow a typical workflow: 
(1) users determine the 2D region through interaction (e.g., painting); (2) a machine automatically constructs the MD region based on the user-defined 2D region; (3) the brushed points are defined as a union or intersection of the set of points within MD and 2D regions. 
Data-driven brushing \cite{martin95vis} allows users to define a 2D region by generating a box that encloses certain areas in the projection, which then generates an MD region as a minimum-size $M$-cube enclosing all the data corresponding to the painted points. In $M$-ball brushing \cite{aupetit14vast}, users can capture MD clusters by defining a circular 2D region; the system then automatically formulates an $M$-ball MD region covering the corresponding data in the MD space. Both approaches bound the MD region to convex shapes, making it hard to discover non-trivial (i.e., any-shaped) clusters. 
Similarity brushing \cite{novotny06wscg} escapes from the problem by allowing users to paint a visual cluster as the 2D region, and defining the MD region as the area covered by the union of $M$-balls centered on the MD data corresponding to the 2D painted points. 

However, all these MD brushing techniques are vulnerable to MDP distortions.  If the 2D brushed region contains FN, the painted data might belong to more than one cluster in the MD space. Moreover, an MD cluster can be split in the projection due to MN, so users will have to brush each of these 2D clusters separately, or even worse, will ignore them if points are spread apart, not forming clear 2D clusters. 

\brush is a data-guided technique that resolves these issues through continuous point relocation.
Instead of keeping continuous 2D and MD regions, we only consider the brushed data points. MN and FN are resolved by pulling them close to or pushing them apart from the currently brushed 2D points, respectively.
In contrast to other techniques, point relocation always generates 2D visual clusters that match MD ones, faithfully representing users' mental model. The method thus works more reliably for the interactive MD cluster analysis.

\section{Design Objectives}

\label{sec:objectives}

Our design objectives tackle the drawbacks of previous brushing techniques for MD data (\hspace{1pt}\oone{} \otwo{} \othree{}\hspace{1pt}) while maintaining their strengths (\hspace{1pt}\ofour{}\hspace{1pt}) (\autoref{sec:dgb}).

\subsubsection*{\oone{} Guide brushing by visually reflecting MD clusters}
Previous data-guided brushing techniques work by converting a 2D brushed region into an MD region.
However, the inconceivability of the MD space makes it difficult for users to understand this conversion, thus lowering the interpretability and controllability of these techniques.
Instead, \brush performs the conversion in the opposite direction: \textit{2D points are relocated to form a visual cluster that reflects an MD cluster}.

\subsubsection*{\otwo{} Allow brushing to be robust against any MDP distortions}

Previous brushing approaches lead to unreliable cluster analysis as they rely on a compact 2D projection region, which is vulnerable to MDP distortions.
In contrast, \brush continuously relocates points to \textit{ensure that 2D neighbors are always true MD neighbors and the MD cluster under focus is not split in the projection}, making the technique work robustly regardless of the type and the amount of distortions.

\subsubsection*{\othree{} Allow the brushing of non-trivial-shaped MD clusters}

In previous brushing techniques \cite{martin95vis, aupetit14vast, ward94vis}, the MD region's shape enclosing brushed points is limited to regular compact domains (hyperspheres or hypercubes).
This limitation makes the techniques hardly support the discovery of clusters with non-trivial shapes typical of real-world MD data.
For example, fitting such a fixed-shaped brush to a non-trivial-shaped cluster can capture out-of-cluster points. 
On the other hand, reducing the brush size to avoid capturing out-of-cluster points may result in missing in-cluster ones, lowering the clustering accuracy. 
In contrast, \brush manages \textit{a discrete set of brushed points instead of compact 2D and MD regions}. This enables users to gradually append new points to the 2D brush corresponding to true MD neighbors of the already brushed points, facilitating the discovery of clusters with arbitrary non-trivial shapes.

\subsubsection*{\ofour{} Minimize the number of hand-tuned hyperparameters}

Previous brushing techniques have at most one hyperparameter that affects brushing results, making them easy to use and learn. 
Similarly, we design \brush to have \textit{a single hyperparameter that affects the granularity of the brushed clusters} and make its value automatically optimized.

\section{\brush}

\brush relocates points within and around the current brushed points in the projection by faithfully reflecting the data distribution around the brushed points in the MD space (\hspace{1pt}\oone{} \otwo{}\hspace{1pt}).
The set of brushed points grows progressively to form a visual cluster that accurately reflects the MD cluster (\hspace{1pt}\othree\hspace{1pt}). By doing so, \brush enables users to conduct a more reliable detailed analysis of MD clusters.

In this section, we first describe the design of \brush following the overall workflow of the technique (\autoref{sec:workflow}).
We then describe additional features of the technique developed for its practical usage in visual analytics (\autoref{sec:additional}).
We describe our technical details in Appendix H.

\subsection{Workflow}

\label{sec:workflow}

\begin{figure*}[ht!]
  \centering
  \includegraphics[width=\textwidth]{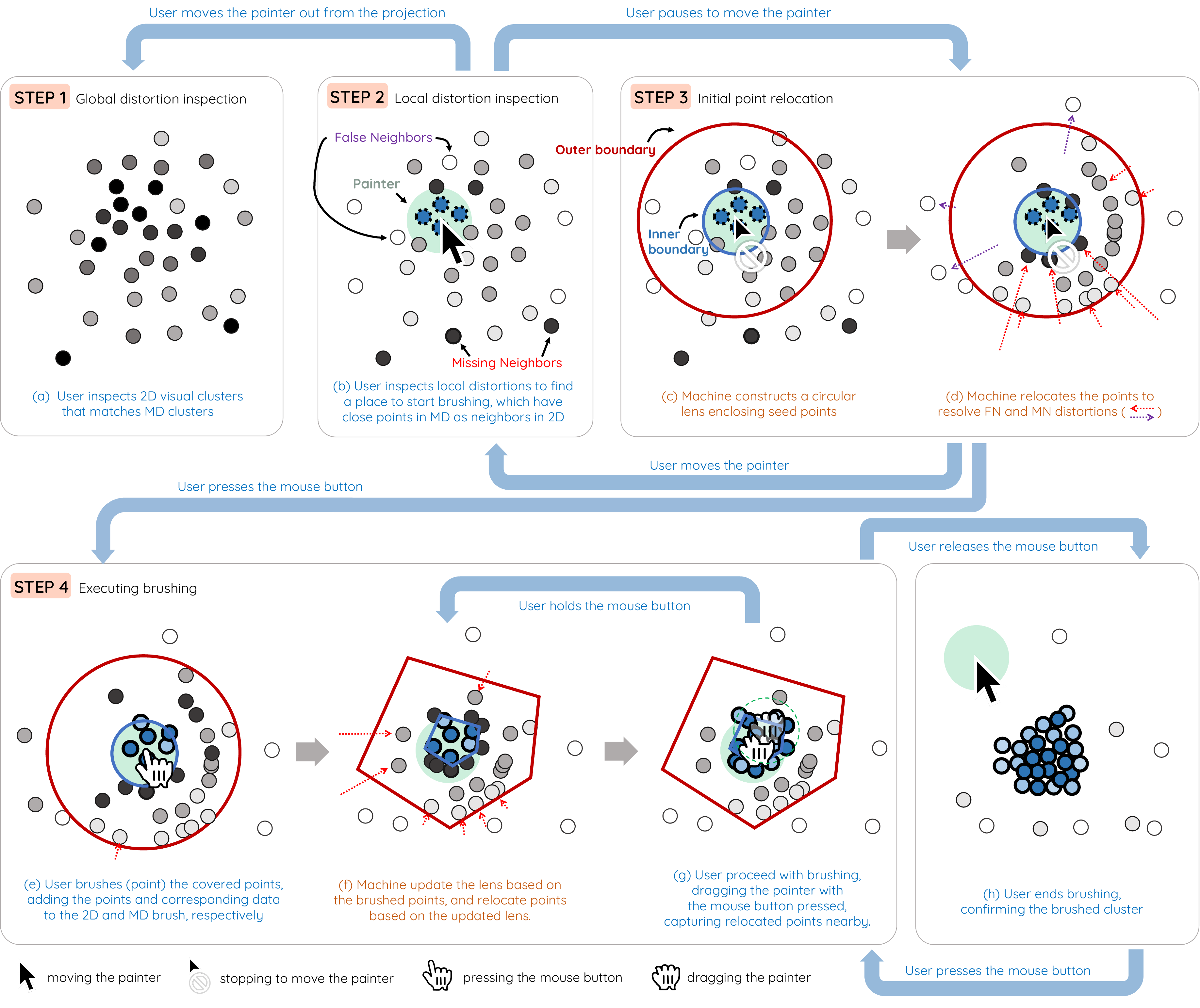}
  \vspace{-7mm}
  \caption{
  Overall workflow of \brush (\ref{sec:workflow}).
  The technique features a lens with inner and outer boundaries depicted as bold blue and red closed lines, respectively.
  Users' actions are explained with blue text and arrows, while the machine's actions are detailed in orange.
  Data points are represented as small circles, i.e., dots, where seed and brushed points are highlighted using thick dotted and solid borders. Seed and brushed points are also highlighted in blue color.
  The opacity of data points depicts MD density in \stepone{} and represents MD closeness to the seed or brushed points in the following steps. 
  }
  \label{fig:workflow}
\end{figure*}


\brush follows a four-step workflow (\autoref{fig:workflow}):
\begin{itemize}[leftmargin=0mm]
    
    \item[]  \stepone{}: Users inspect how the MD data distribution matches the 2D visual clusters to decide the best places to initiate brushing. 

    \item[] \steptwo{}: Users inspect local MN and FN distortions around these candidate places by hovering the painter over the points.

    \item[] \stepthree{}: A pause of mouse move initiates a transient relocation of the points, correcting the local distortions. 

    \item[]  \stepfour{}: Users execute brushing by dragging the painter while the mouse button is pressed, progressively capturing the covered points.
Lens construction and point relocation are performed iteratively based on the current 2D location of brushed points and the MD distribution in their vicinity.
\end{itemize}


\subsubsection*{\stepone{} Inspecting global distortions}

For reliable MD cluster analysis, brushing shall ideally initiate at a place close to the core of an actual MD cluster. To support the task, we encode each point as a snippet representing the corresponding MD datum (\autoref{fig:teaser}). For example, each point in image datasets can be represented as an image snippet. Also, points in tabular datasets can be represented using glyphs \cite{kammer20tvcg}, e.g., aster plots \cite{kwon17tvcg}. By visualizing snippets, users can spot 2D visual clusters (high mutual proximity) matching with MD clusters (high similarity between snippets) (\hspace{1pt}\oone{} \otwo{}\hspace{1pt}), recognizing these locations as good candidates for initiating brushing.

We further aid this step by encoding the MD density of the data in the MD space through the opacity of the corresponding points or snippets (\autoref{fig:workflow} (a)). 
Based on density encoding, users can check the trustworthiness of a 2D visual cluster by comparing whether the 2D density, represented by proximity between points, matches MD density.
A visual cluster with higher 2D and MND density than other visual clusters can be considered a good candidate. 
The best location to start the brushing is the high-density central part of the 2D visual cluster, which corresponds to the core region of the MD cluster.
Other cases showing a density mismatch reflect MDP distortions and should be avoided. More comprehensive distortion visualizations could be used (e.g., visualizing FN and MN distortions \cite{lespinats11cgf, jeon21tvcg}), but we preferred visualizing the density to make users easily learn the technique.


\boldsubsubsection{MD density}
We define the MD density of a point $p$ as \dens{p} $= \sum_{q \in P}$ \simil{p}{q}, following Density-peak clustering \cite{rodriguez14science, liu18infosci}, where \simil{p}{q} represents the similarity between points $p$ and $q$, and $P$ denotes the entire set of points in the data.

\boldsubsubsection{MD similarity measure}
We use the Shared-Nearest Neighbors (SNN) similarity \cite{ertoz02siam}, which assigns higher similarity to the pairs of points sharing more $k$-Nearest Neighbors ($k$NN). Formally, the SNN similarity between $p$ and $q$ is defined as \texttt{sim}$_k(p, q) = \sum_{(m,n) \in S_{p, q}} (k + 1 -m) \cdot (k + 1 -n)$; $S_{p,q}$ represents a set containing pairs $(m, n)$ fulfilling $p_m = q_n$ where $p_i$ denotes an $i$-th nearest neighbor of $p$ and $q_i$ denotes an $i$-th nearest neighbor of $q$. 
We select SNN as this metric is shift-invariant \cite{lee11pcs}, making it alleviate the curse of dimensionality \cite{lee11pcs, jeon22arxiv} and thus better represents the cluster structure of MD spaces compared to other metrics (e.g., $k$NN, Euclidean distance) \cite{ertoz02siam, liu18infosci, jeon21tvcg}. 
We fix $k$ as the square root of the number of points, following the recommendation by Chaudhuri and Dasgupta \cite{chaudhuri14neurips}.

\begin{figure*}[t]
  \centering
  \includegraphics[width=\linewidth]{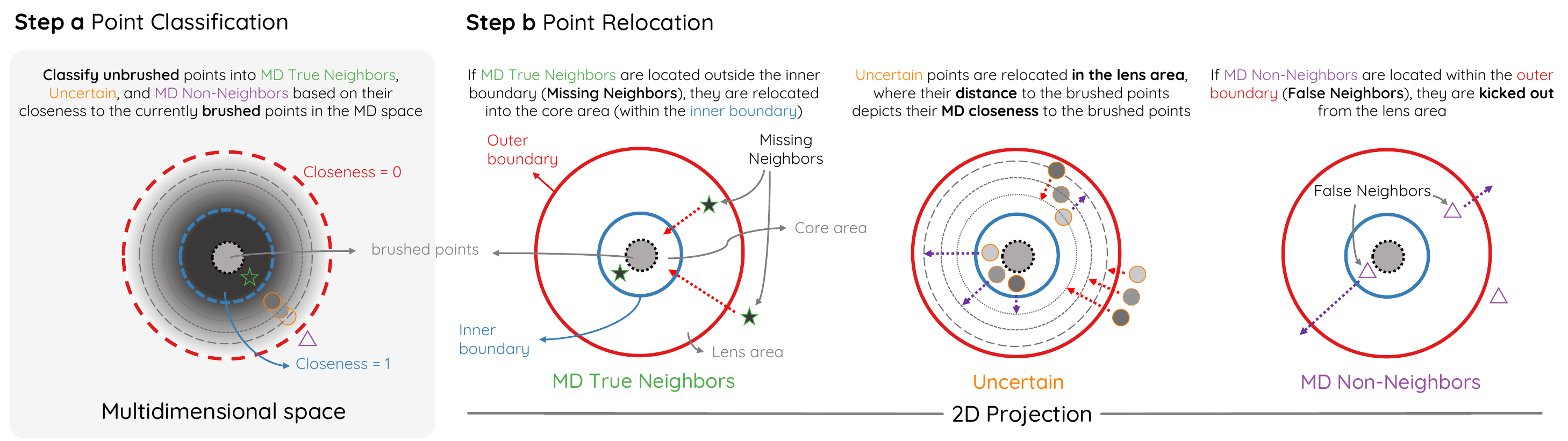}
  \caption{
    Illustration on how \brush relocates points in \stepthree{} and \stepfour{}. The machine first examines the MD closeness of unbrushed points to the brushed points (or seed points in \stepthree{}) (3-a), then relocates those points in the projection to reflect that MD closeness.
  }
  \label{fig:relo_cases}
\end{figure*}

\subsubsection*{\steptwo{} Inspecting local distortions}

In this step, users can ``skim'' local distortion of the projection by moving the painter over the points (\autoref{fig:workflow} (b)). This is done by (1) finding \textit{seed points} within a painter, then (2) visualizing the MD closeness of any point to the seed points \cite{aupetit07neurocomputing}. 
Compared to \stepone{}, this step provides a more stringent inspection of local distortions that can help users locate the best candidate for initiating brushing. 

The detailed procedure is as follows. First, the machine determines the seed points as condensed points the painter covers.
This is done by identifying the covered point with the highest MD density and only using its close neighbors as seed points (see \textit{Finding seed points} below).
It is important to avoid using every point covered by the painter as seed points because they may consist of points coming from two or more distinct MD clusters due to FN; if this happens, new points brushed from these distinct MD clusters will erroneously agglomerate into a single visual cluster (\hspace{1pt}\oone{} \otwo{}\hspace{1pt}). 
The seed points are then highlighted with a color corresponding to the current brush.

Then, the remaining points' MD closeness to the seed points is encoded as their opacity (i.e.,  closer points are darker). This graphical encoding informs users' brushing decisions by indicating more reliable locations to start brushing (\hspace{1pt}\oone{}\hspace{1pt}). Users can identify FN in the projection as points with bright or non-highlighted markers near the seed points and MN as points with dark markers far from the seed points. 

\boldsubsubsection{Finding seed points}
Constructing a set of seed points starts by identifying the initial seed point $p_{initial}$ with the highest MD density among the subset of points $C$ covered by the painter: $p_{initial} = \argmax_{p \in C} \texttt{dens}(p)$.
Then, a set of seed points is defined as the $\kappa$ nearest neighbors ($\kappa$NN) of $p_{initial}$ based on SNN similarity in the MD space, 
where the machine automatically sets $\kappa$ as the maximum value such that all $\kappa$NN are still covered by the painter (\hspace{1pt}\ofour\hspace{1pt}). Users can adjust the size of the painter and so $\kappa$, to make MD brushing more condensed or relaxed. By this definition, the seed points and the initial brush cannot contain FN. 

\boldsubsubsection{MD closeness}
A closeness between a point $p$ and a set of points $C$ is: \close{p}{C} $=\sum_{q \in (\kappa NN \cap C)} \texttt{sim}_k(q, p) / \sum_{q \in \kappa NN} \texttt{sim}_k(q, p)$.
The more $\kappa$NN of $p$ are members of the cluster $C$, the closer $p$ is to $C$. Moreover, by definition, the seed points of the initial brush have the maximum closeness.
We do not naively average the similarity of $p$ to the points within $C$ because this will make the closeness depend on cluster characteristics like density or size.

\subsubsection*{\stepthree{} Initiating point relocation}
Though global (\hspace{1pt}\stepone\hspace{1pt}) and local distortion (\hspace{1pt}\steptwo\hspace{1pt})  inspections support users in finding a good candidate region, the points nearby likely suffer from MDP distortions, not accurately reflecting the local MD data distribution. We thus provide a \textit{point relocation} process that corrects the distortions relative to the current brush (initially, the seed points covered by the painter)  (\autoref{fig:workflow} (c-d)).
Inspired by the Proxilens approach  \cite{heulot13vamp}, users can trigger transient point relocation by halting the painter for a short time, which is determined as 800ms in our implementation through an iterative design process. Correcting for local distortions can be viewed as ``jumping'' into the MD space, as it makes the 2D distribution around the painter and the currently brushed points better reflect the local MD data distribution. Users can reverse the current point relocation (jumping back to the 2D space) by moving the painter again. 

To perform relocation, the system first constructs a magic lens around seed points in the projection  (\autoref{fig:workflow} (c)). The lens consists of (1) a core area delimited by an inner boundary that tightly encloses seed points and (2) an annulus lens area around the core, enclosed between the inner boundary and an outer boundary (\autoref{fig:relo_cases}, blue and red solid circular boundaries, respectively). 



\definecolor{mymygreen}{RGB}{77,175,74}
\definecolor{mymyorange}{RGB}{255,127,0}
\definecolor{mymypurple}{RGB}{152,78,164}

\def\mdtrueneighbors{{\color{mymygreen}{\textit{MD True Neighbors}}}\xspace}

\def\mdnonneighbors{{\color{mymypurple}{\textit{MD Non-Neighbors}}}\xspace}
\def\uncertain{{\color{mymyorange}{\textit{Uncertain}}}\xspace}

Afterward, point relocation is performed according to the lenses so that points distribution around seed points in the projection can reflect the distribution around the seed points in the MD space(\hspace{1pt}\oone{}\hspace{1pt}) (\autoref{fig:workflow} (d)). 
The relocation of a point thus depends on its MD closeness to the seed points (\autoref{fig:relo_cases}). 
If the closeness is 1, the point is considered as \mdtrueneighbors, which means that the point belongs to the core MD cluster formed by the seed points.
If the closeness is 0, the point is categorized with \mdnonneighbors, denoting that it is far apart from the seed points in the MD space.
If the value is between 0 and 1, the point is considered to be \uncertain,  forming a ``fuzzy'' neighborhood in the MD space.
While the machine relocates \mdtrueneighbors (\textit{i.e} MNs) into the inner boundary, \mdnonneighbors (\textit{i.e.} FNs) are repelled from the lens area outside the outer boundary, and \uncertain points are relocated within the lens area, close to the inner boundary, in proportion to their MD closeness to the core MD cluster. 
Points are relocated to the correct position with an animated transition. 
Relocation corrects MDP distortions, making points in the inner lens correspond to True Neighbors, as FNs are repelled from the lens and MNs attracted within the lens (\hspace{1pt}\oone{}\hspace{1pt}).

It is worth noting that the \uncertain points and their interpolated relocation are a crucial part of \brush, as it is up to users to decide whether a point will be brushed or not. The \uncertain points falling into the outer lens are natural candidates for brushing; they are also geometrically the next ones that can be captured by the painter (\hspace{1pt}\oone{} \otwo{}\hspace{1pt}) (see \textit{Dynamical update} in \stepfour{}). 

\boldsubsubsection{Initial lens construction}
The initial inner and outer lens boundaries are defined as circular boundaries centered on the painter.
We set $\tau$ as both the inner boundary's radius and the radius of the painter.
By doing so, the painter covers both seed points and \textit{MD True Neighbors}, forcing them to be brushed when users execute brushing (\hspace{1pt}\stepfour{}\hspace{1pt}). 
We also define the radius of the outer boundary as $3\tau$, setting the lens area's width as the painter's diameter (i.e., $2\tau$). 
We justify this decision while describing how we construct the outer lens boundary in \stepfour{} (\textit{Outer boundary construction}).

\subsubsection*{\stepfour{} Executing brushing}

Once the transient relocation is settled, users can initiate brushing by pressing the mouse button (\autoref{fig:workflow} (e)).
The machine appends the points covered by the painter (which naturally contains seed points) to the brushed set of points. The brushed points are highlighted with the color corresponding to the brush. 
Then, the machine updates the lens and relocates the remaining points based on the new set of brushed points (\autoref{fig:workflow} (f)).
If users drag the painter while keeping the mouse button pressed (\autoref{fig:workflow} (g)), the brushed points, the painter, and the lens are updated accordingly, and the relocation takes place again in a continuous cycle. 
Such gradual updates enable users to agglomerate new brushed points in an arbitrary direction in the MD space, thus allowing the brushing of an MD cluster with an arbitrary shape (O3).

If users decide to end brushing and confirm the brushed MD cluster, they can end the cycle by releasing the mouse button.
Such decisions can be made when (1) there are no more unbrushed image snippets (i.e., data points) that look similar to the ones inside the set of brushed snippets or (2) newly brushed points have a relatively lower density than the previously added points. Auxiliary visualizations (e.g., parallel coordinates plot or heatmap) can also guide users in deciding the boundary of brushed clusters (\autoref{sec:scenario}).
Here, users can again go back to brushing by pressing the mouse button or erase points that are not intended to be in the brush (\autoref{sec:erasing}).

\revise{
Note that by allowing users to make the final decision, our technique can effectively handle noisy clusters, i.e., semantic clusters that are not well separated in the data space. For example, \brush can be used to highlight image snippets that are visually distinguishable to humans but indistinguishable by a distance metric. We demonstrate the effectiveness of this feature in \autoref{sec:usecaselabeling}.}

During brushing, the brushed points are also relocated to aid interaction. 
First, we uniformize their locations.
The uniformization removes empty space within the inner boundary, making the 2D proximity between each unbrushed point and the brushed points accurately reflect their closeness. 
It also removes overlap between the points, supporting users in visually investigating snippets. 
Then, the points are successively relocated to better reflect the original data distribution of the MD cluster.
The points in which corresponding data have high MD closeness to the brushed points move near the center of the core area of the lens, and the ones with low closeness move near the boundary of the inner lens. 
We achieved this by swapping the positions of the points to align their distances to the inner boundary with closeness.
This makes the 2D visual cluster a better match with the MD cluster (\hspace{1pt}\oone{}\hspace{1pt}) and helps users readily erase points with low closeness (\hspace{1pt}\otwo\hspace{1pt}).

\boldsubsubsection{Inner boundary construction}
While brushing, the inner boundary is set as a convex hull enclosing the brushed points. 
We use a convex hull as it is computationally cheap ($O(n\log n)$ for $n$ points) while tightly enclosing the brushed points compared to alternatives (e.g., boundary circle) and has no hyperparameter to tune (\hspace{1pt}\ofour\hspace{1pt}). 

\boldsubsubsection{Outer boundary construction}
The outer boundary is constructed by offsetting each corner of the inner boundary to maintain the width of the lens area. Thus, points with the same MD closeness to brushed points are displayed at the same distance to the core lens. 
This approach also makes the relocation of points isotropic, not biased by the direction from which they originate. It supports our encoding where visual proximity matters while provenance direction has no specific importance. We detail this justification and design alternatives in Appendix D.

As mentioned in \stepthree{}, we use $2\tau$ offset to match the lens area's width with the painter's diameter. Thus, when an \uncertain point enters the painter, lying on its circular edge, its MD closeness to the brushed points corresponds roughly to the proportion of the painter area overlapping the core lens, a visual indicator easy to estimate. 
For example, if the painter fully overlaps with the core lens, any point within the painter can belong to the brush but none within the lens area (i.e., the accepted closeness is only $\alpha = 1$).
If the painter $\alpha$-overlap with the core lens, points with MD closeness above $\alpha$ can belong to the brush (the accepted closeness range is $[\alpha, 1]$).
Finally, if the painter stays entirely within the lens area, outside the core lens, touching the outer boundary,  all the points covered by the painter can belong to the brush (the accepted closeness range is total: $[0,1]$). 

\boldsubsubsection{Dynamical update}
As the brush is dynamically updated with the \uncertain points captured by the painter, the brush grows up on the painter's side at a certain speed due to relocation animated transitions and \textbf{core lens uniformization} (see below). This increases the overlap area of the core lens with the painter, hence the acceptance threshold $\alpha$, and reduces the painter's covering of the lens area, preventing further \uncertain points from being captured and a positive feedback loop that would lead to a loss of control. Users must keep moving the painter toward the nearby lens area to lower the acceptance threshold back to the desired value, capturing new \uncertain points down to that level. Thus, users intending to brush MD data carefully will naturally move painter slower, thus maintaining a higher $\alpha$ and a more stringent acceptance of \uncertain points as \mdtrueneighbors.

\boldsubsubsection{Core lens uniformization}
We used the centroidal Voronoi tessellation \cite{du99siam} based on Lloyd's algorithm \cite{lloyd82tit} to uniformize the distribution of brushed points. We clip the Voronoi cells to fit the inner boundary so that the points remain within the core lens.

\subsection{Features for Enhanced Usability}

\label{sec:additional}

We discuss features of \brush that improve its usability and applicability in cluster analysis.

\subsubsection{Erasing Mode}

\label{sec:erasing}

Users can erase points that are not intended to be brushed using erase mode. 
The mode is enabled when (1) brushing is paused by releasing the mouse button during \stepfour{} and (2) the keyboard shift button is pressed. When enabled, users can erase points by hovering the painter over the points while keeping the mouse button pressed. The painter's color becomes red.

Natural candidates for erasing are points with low MD closeness to the brushed points.
As the system pushes such points with low closeness near the inner boundary (\hspace{1pt}\stepfour{}\hspace{1pt}), users can readily erase them by moving the painter in the lens area of the brush slightly overlapping the core lens to capture these outliers.

The erased points are then relocated like other unbrushed points based on their closeness to the updated brush.


\subsubsection{Multiple Brushes}

In previous brushing techniques, multiple brushes are
often provided to compare multiple sets of points, widening the analytic search space \cite{aupetit14vast, ward94vis}.
We also allow users to control multiple brushes while
distinguishing them with different colors. Users can pause the current brush and switch the focus to another brush by pressing a button with the same brush color. We showcase the utility of multiple brushes in our use case (\autoref{sec:scenario}).

\subsubsection{Contextualization within the Original Projection}

\label{sec:contextualization}

The point relocation mechanism corrects distortions around the brushed points (\hspace{1pt}\otwo{}\hspace{1pt}) but can introduce new distortions elsewhere in the MDP, making it difficult for users to understand the brushed cluster in the context of the original MDP. 
\revise{Moreover, the relocation may reinforce confirmation bias, leading 
users to preferentially brush points that resemble those already brushed.}
To mitigate such side effects, we allow users to restore all points to their original positions through an animated transition while preserving their color to indicate brush identity.
\revise{
This feature helps users contextualize brushing results within the original MDP, reducing selection bias introduced by the current brushing and encouraging the exploration of alternative analytical paths.
Our use cases (\autoref{sec:scenario}, \ref{sec:usecaselabeling}) show the practical benefits of this contextualization in practical scenarios.
Future work could explore history management for clusters \cite{menin21infovis}, which would support  hypothesis-driven analysis and enhance navigation across different brushing results---an improvement beneficial to all brushing techniques.
}

\section{Evaluating the Robustness of\\Distortion-Aware Brushing}

\label{sec:userstudy}

We conduct two controlled user studies to validate the effectiveness of \brush.
The studies investigate how \brush's performance in extracting MD clusters (accuracy and task completion time) is affected by the amount of distortions (O2; \autoref{sec:comparisonstudy}) and the non-triviality of the shape of MD clusters (O3; \autoref{sec:detailstudy}). 
Both studies compare \brush against three existing brushing techniques for MDP.
We do not empirically investigate the satisfaction of O1 and O4, as they are fulfilled by design (\autoref{sec:workflow}).
%


\subsection{User Study 1: Robustness Against Distortions}

\label{sec:comparisonstudy}

\begin{figure*}
    \centering
    \includegraphics[width=\linewidth]{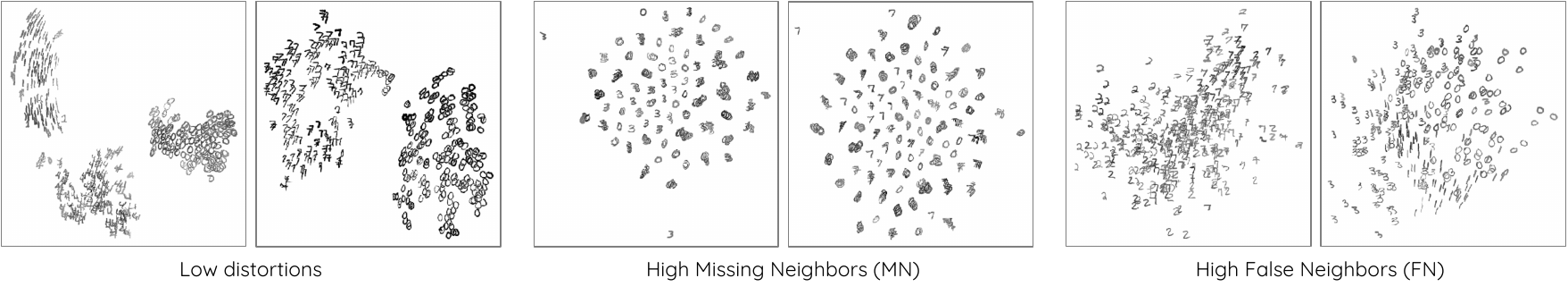}
    \vspace{-7mm}
    \caption{The example projections (i.e., stimuli) used in our experiments with different amounts of distortions. In study 1 (\autoref{sec:comparisonstudy}, we treat the amount of distortions as an independent variable, namely \textsc{DistortionAmount}. In study 2 (\autoref{sec:detailstudy}), it is controlled as a confounding variable. }
    \label{fig:stimuli}
\end{figure*}

\subsubsection{Objectives and Design}

\label{sec:comparisonstudydesign}

We aim to compare \brush with baselines regarding accuracy in capturing MD clusters and their robustness to the amount of DR distortions.
We also want to assess how global  (\hspace{1pt}\stepone\hspace{1pt}) and local (\hspace{1pt}\steptwo\hspace{1pt}) distortion inspections (\autoref{sec:workflow}) affect user performance of the brushing techniques.
To achieve such goals, we design the study to have three independent variables:
\begin{itemize}
    \item Amount and type of distortions (\textsc{DistortionAmount})
    \begin{itemize}
        \item \textit{Low distortion}, \textit{High MN}, and \textit{High FN}
    \end{itemize}
    \item MD brushing techniques (\textsc{Techniques})
    \begin{itemize}
        \item Three baselines (\textit{Data-driven brushing} \cite{martin95vis}, \textit{$M$-Ball Brushing} \cite{aupetit14vast}, \textit{Similarity brushing} \cite{novotny06wscg}) and \textit{\brush}.
    \end{itemize}
    \item Availability of global  and local distortion inspections (\textsc{DistortionInspection})
    \begin{itemize}
        \item \textit{No inspection}, \textit{Only global}, and \textit{Both global and local}
    \end{itemize}
\end{itemize}
resulting in a within-subject experiment with 3 \textsc{[DistortionAmount]} $\times$ 4 \textsc{[Techniques]} $\times$ 3 \textsc{[DistortionInspection]} $=36$ trials per participant.
Note that controlling \textsc{DistortionInspection} ensures the fairness of our experiment in comparing \brush and baseline techniques.
This is because baseline techniques' task performance may also benefit from the inspection functionalities.
The detailed study design is presented in the following subsections:

\boldsubsubsection{\revise{Baseline brushing techniques}}
\revise{We select data-guided brushing techniques (\autoref{sec:dgb}) as baselines because they share a common analytic goal with \brush: identifying MD clusters. Axis-guided brushing (\autoref{sec:axisguided}) techniques are not included because they (1) operate on multiple coordinated attribute-based views (SPLOMs) instead of a single MDP and (2) serve a different analytic purpose (\autoref{sec:axisguided}), limiting their capability to identify MD clusters (see Appendix I). 
}


\boldsubsubsection{Task}
We ask participants to perform interactive labeling \cite{sacha17tvcg, peltonen13eurovis}. The aim of the task is to label a single designated cluster in the MD data by brushing them on the 2D MDP represented by a monochrome scatterplot. 
We pick the task as it is widely used to explore MD data in visual analytics \cite{sacha17tvcg, peltonen13eurovis, jiazhi21tvcg}, and is an important use case of previous brushing techniques \cite{aupetit14vast, novotny06wscg}.

\boldsubsubsection{Procedure}
One experimenter manages the experiment for all participants individually and in person.
After a participant signs the consent form, the experimenter explains the concept of MD data and why 2D projections cannot precisely depict them. Then, the experimenter details the tasks and goals of the experiment. During the introduction, the participants are free to ask questions.

Participants are then exposed to 36 trials; each is associated with a single combination of the independent variables.
We divide the trials into four sessions, where each session is assigned to a single \textsc{Technique}.
In each session, after the experimenter demonstrates the technique, participants are given a maximum of five minutes to practice and pose questions.
For the practice, we use a dataset and projections different from those used for the main study (detailed below).
The order of the sessions (i.e., the order of the \textsc{Techniques}) is counterbalanced using a four-level Latin square design (Appendix F).
Each of these sessions contains nine trials (every combination of \textsc{DistortionAmount} and \textsc{GlobalInspection}), where the order of these trials is again counterbalanced using a Latin rectangular design (Appendix F). As we notice that each trial takes at most around 120 seconds in a pilot study, we set no specific time limit for each trial. 

Finally, we conducted a semi-structured post-study interview investigating the usability of brushing techniques and task difficulty (Appendix A). The experiment took less than 50 minutes for all participants. We detail our apparatus in Appendix J.

\boldsubsubsection{Measurements}
We record labeling results and task completion time of each trial. 
We record the task completion time as the duration from the first mouse hover to the stimuli to the click of the ``finish trial’’ button.
We also convert labeling results into F1 accuracy scores with respect to the ground-truth clusters.


\boldsubsubsection{MD dataset and ground truth clusters}
We use the MNIST dataset as a reference MD dataset to generate projections that will be used as scatterplot stimuli.
We reduce the dimensionality of the MNIST dataset to 10D using PCA \cite{pearson01pmjs} to prevent previous brushing techniques that rely on convex-shaped MD region (Data-driven brushing, $M$-ball brushing) 
 to suffer from the curse of dimensionality \cite{bellman66science} (see Appendix B), making our experiment fair.
 
We consider each class (i.e., digit) as a potential ground truth cluster. 
We thus render points as snippet images (i.e., digits) so that participants can visually distinguish different digits and readily determine the boundary of brushing.

However, classes may not be well separated in the original MD space \cite{jeon24tvcg2, aupetit14beliv}. 
Therefore, we first identify class pairs with high separability following Aupetit \cite{aupetit14beliv} to create valid ground-truth clusters and allow reliable evaluation. 
We first compute the separability of class pairs over 96 labeled MD datasets available using the between-dataset Calinski-Harabasz ($CH_{btwn}$) Index \cite{jeon22arxiv}, then pick the index value corresponding to the 90th percentile as a separability threshold.
We use $CH_{btwn}$ as the index is proven to work robustly and fairly regardless of the dimensionality \cite{jeon24tvcg2, jeon22arxiv}
Then, for each stimulus, we select pairs of classes in the MNIST dataset that are mutually separable with a $CH_{btwn}$ separability index higher than the separability threshold.





\boldsubsubsection{Confounding variables}
We identify and control three confounding variables: the number of points of each cluster, the number of clusters, and the non-triviality of the shape of the cluster designated to be brushed.
To control the non-triviality of a cluster, we first fit the Gaussian Mixture model with a single Gaussian distribution to the cluster in the MD space.
We then use the maximum log-likelihood score, quantifying how well the Gaussian fits the data distribution as a proxy for non-triviality. 
We identify the top three classes with high non-triviality and the bottom three classes with low non-triviality of the MNIST dataset as \textit{High} and \textit{Low} non-triviality clusters, respectively, designating the remaining four classes as \textit{Intermediate} clusters. 
We also prepare three different settings for the number of points (\textit{100}, \textit{150}, and \textit{200}) and clusters (\textit{two, three,} and \textit{four}). 
The total number of combinations of the confounding variables is $3 \times 3 \times 3 = 27$, but we reduce it to nine using a three-level Latin square design. 
\reviset{Note that the number of points thus varies from 200 (100 points $\times$ two clusters) to 800 (200 points $\times$ four clusters).}
We maximally equalized the assignment of these nine combinations to all participants (12) and trials (36) using the Latin rectangle design (See Appendix F for the detailed assignment).

\boldsubsubsection{Stimuli (i.e., MDPs) generation}
We have nine combinations of confounding factors and three different \textsc{DistortionAmount} settings; we thus need $9 \times 3 = 27$ types of stimuli. 
When a trial requires a stimulus with $\mathcal{N}$ clusters, $\mathcal{M}$ points per cluster, $\mathcal{L}$ non-triviality, and $\mathcal{O}$ \textsc{DistortionAmount}, we first randomly sample $\mathcal{N}$ clusters while ensuring that at least one of them has $\mathcal{L}$ non-triviality. We then randomly sample data points to make each cluster have $\mathcal{M}$ points and generate a projection having $\mathcal{O}$ \textsc{DistortionAmount} (detailed below) as a final stimulus. One of the clusters with $\mathcal{L}$ non-triviality is randomly designated to be brushed.
As we use the MNIST dataset, we designate the cluster by informing the corresponding class. 
All stimuli are precomputed before the experiment. Please refer to \autoref{fig:stimuli} for the example projections we use.

\boldsubsubsection{Levels of \textsc{DistortionAmount}}
We generate projections with \textit{Low distortion} using $t$-SNE \cite{maaten08jmlr}. We optimize two hyperparameters that significantly impact the projection results, perplexity, and learning rate \cite{gove22vi}. We use Bayesian optimization \cite{snoek12nips} while using the F1 score of Trustworthiness \& Continuity (T\&C) \cite{venna06nn} as the target function.
We use T\&C as they are widely used metrics for detecting FN and MN \cite{jeon21tvcg, jeon24tvcg2, nonato19tvcg, jeon23vis}, respectively.
We also create projections with \textit{High FN} using the random orthogonal projection technique. As the random orthogonal projection always decreases the distances between points, we can expect the resulting projection to have relatively high FN but relatively low MN.
We create \textit{High MN} projections by executing $t$-SNE with an extremely low perplexity setting of 1. Given that low perplexity values cause $t$-SNE to prioritize local structure, setting the perplexity this low can result in the algorithm missing close neighbors and splitting cluster apart, thereby leading to more MN \cite{wattenberg2016tsnetuning}.
We verify the validity of our settings in Appendix E by reporting the T\&C scores of the resulting projections.

\boldsubsubsection{Training session for each brushing technique}
We use PCA projection of a randomly chosen set of three highly separated classes for the training. We use PCA as it is a very standard projection technique that generates FN distortions but less than random orthogonal projections. Therefore, using PCA, participants can experience each brushing technique's capability to address distortions without being frustrated by the task's difficulty. We turn on both global and local inspection during the training session. We inform participants that these functionalities can be turned off during the main experiment.

\boldsubsubsection{Participants}
We recruited 12 participants (identified as E1P1--E1P12)
from a local university (nine males and three females, aged 22--32 [26.4 $\pm$ 3.2]).
All participants had undergraduate-level experience in statistics or linear algebra and thus could readily understand the concept of MD data and brushing techniques. 
They also had experience in using standard brushing techniques like lasso or box selections.
We compensated each participant with the equivalent of  US \$10 for their participation.

\begin{figure*}
    \centering
    \includegraphics[width=\linewidth]{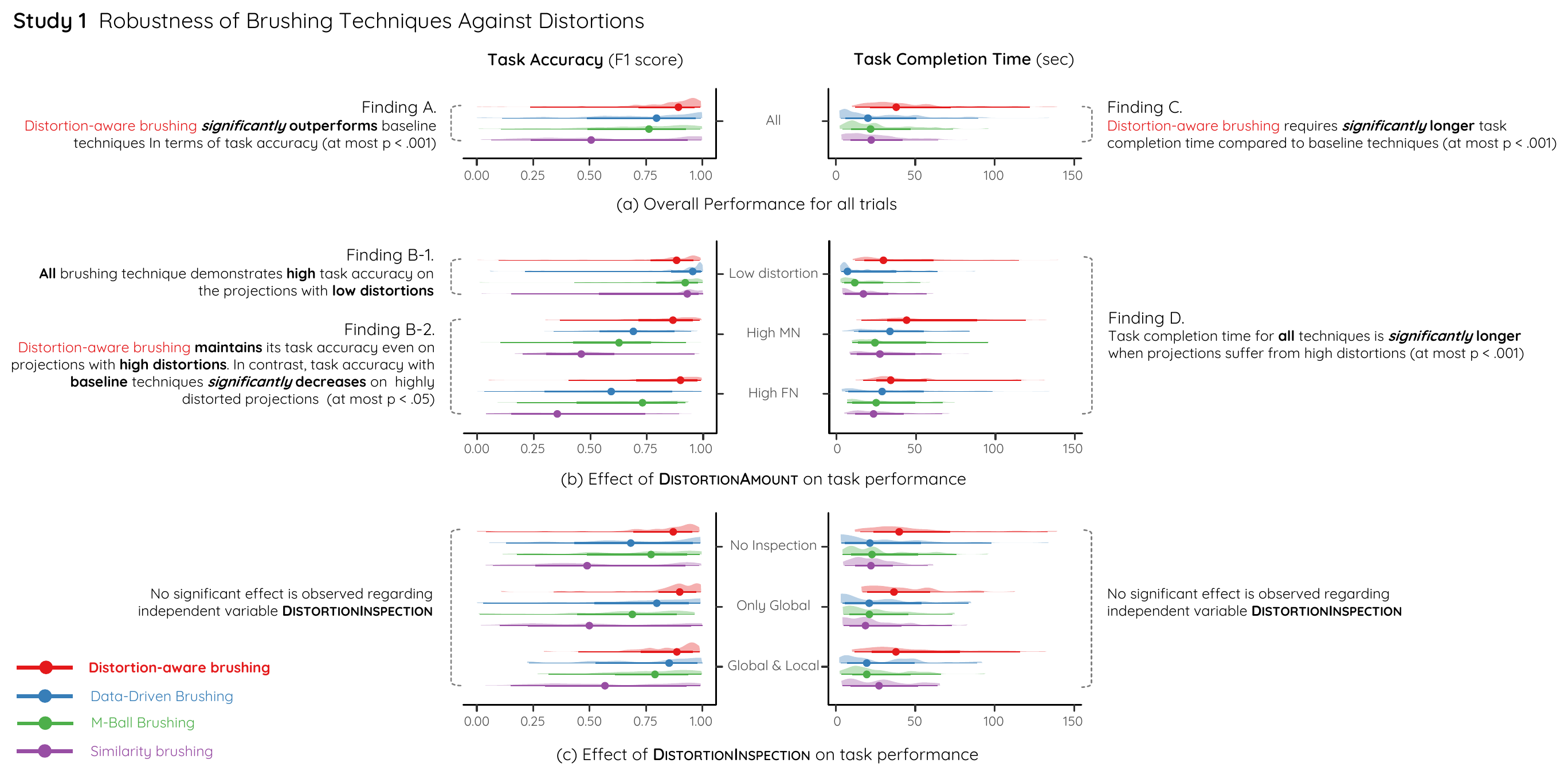}
    \vspace{-6mm}
    \caption{Study 1 results demonstrating the robustness of brushing techniques against the amount of distortions. Overall, \brush significantly outperforms baseline techniques in terms of task completion time, but requires participants more time to complete the task. Please refer to \autoref{sec:study1results} and \autoref{sec:study1discuss} for detailed analysis results and discussions, respectively.}
    \label{fig:exp1_results}
\end{figure*}

\subsubsection{Quantitative Results}

\label{sec:study1results}

We detail our study findings, which are also depicted in \autoref{fig:exp1_results}.

\boldsubsubsection{Analysis on overall task accuracy}
Aligned with the main study objective (\autoref{sec:comparisonstudydesign}), we investigate how three independent variables (\textsc{DistortionAmount}, \textsc{DistortionInspection}, and \textsc{Techniques}) affect clustering accuracy, i.e., F1 score (\autoref{fig:exp1_results} left).
We execute the three-way repeated measure ANOVA (RM ANOVA) with Bonferroni correction for post-hoc analysis. 

As a result, we find significant main effect on \textsc{Techniques} ($F_{3, 33} = 36.361$, $p < .001$) and \textsc{DistortionAmount} ($F_{2, 22} = 17.102$, $p < .001$). For \textsc{Techniques}, post-hoc analysis reveals that \brush shows significantly higher accuracy compared to baseline techniques ($p <.01$ for Data-driven brushing case, $p < .001$ for other cases). There was no other significant difference.
This leads to our first finding:

\finding{A}{\brush is, overall, the most accurate technique among all competitors for selecting clusters in MDPs.}

\noindent
In terms of \textsc{DistortionAmount}, we find that task accuracy is significantly better on the projections with low distortions compared to the ones with high MN or high FN ($p < .001$ for both cases), which is straightforward.

\boldsubsubsection{Interaction effect analysis on task accuracy}
 Using RM ANOVA, we also find a significant interaction effect between \textsc{Techniques} and \textsc{DistortionAmount} ($F_{6, 66} = 7.272$, $p < .001$).  To deep dive into this interaction, we divide the trials into three groups with different \textsc{DistortionAmount} (\textit{Low distortion}, \textit{High MN}, and \textit{High FN}). 
We then conduct additional one-way ANOVA examining how \textsc{Techniques} affect the F1 accuracy scores of each group. 

As a result, we find that \textsc{Techniques} variable significantly influences the accuracy with the projections having \textit{High MN} ($F_{3, 140} = 20.764$, $p < .001$) or \textit{High FN} ($F_{3, 140} = 19.318$, $p < .001$). However, it has no significant effect for the case with \textit{Low distortions} ($F_{3, 140} = 0.847$, $p = .470$), where all techniques show substantially high performance (over 0.85 in average; \autoref{fig:exp1_results}b left).
 Post-hoc analysis shows that if projections have  \textit{High MN} or \textit{High FN}, \brush significantly outperforms baseline techniques (\textit{High MN}: $p < .01$ for Data-driven brushing case, $p <  .001$ for other cases; \textit{High FN}: $p < .05$ for $M$-Ball brushing case, $p < .001$ for other cases). 
 \autoref{fig:exp1_results}b (right) shows that such change occurs as the accuracy of \brush stays still while the accuracy of other techniques decreases. The results can be summarized as follows:

\finding{B-1}{Every brushing technique we considered is accurate for the MDPs with low distortions.}

\vspace{-2.5mm}
\finding{B-2}{\brush maintains its high task accuracy on projections with high MN and FN. On the other hand, baseline techniques work poorly on these projections, having significantly lower task accuracy compared to \brush.}

No other significant interaction effect has been found. 

\boldsubsubsection{Analysis on overall task completion time}
We investigate how three independent variables affect task completion time  (\autoref{fig:exp1_results} right). As with previous analysis on task accuracy, we run the three-way RM ANOVA for that purpose, and use Bonferroni correction for post-hoc analysis.

The analysis reveals significant main effect on \textsc{Techniques} ($F_{3, 33} =8.557$, $p <.001$) and \textsc{DistortionAmount} ($F_{2, 22} =7.162$, $p <.01$). For \textsc{Techniques}, post-hoc analysis shows that \brush shows significantly longer task completion time compared to baseline techniques ($p < .001$ for all cases). This summarizes into:

\finding{C}{\brush requires significantly longer task completion
time than baseline techniques.}

In terms of \textsc{DistortionAmount}, post-hoc analysis reveals that the techniques require longer completion time for the projections with \textit{High MN} and \textit{High FN} compared to the \textit{Low distortions} cases. As no interaction effect coupled with \textsc{DistortionAmount} exists, the results lead us to the following finding:

\finding{D}{Regardless of the type of brushing techniques, the task completion time was significantly
longer when there exist distortions in MDPs.
}

\boldsubsubsection{Interaction effect analysis on task completion time}
RM ANOVA identifies significant interaction effect on \textsc{DistortionAmount} and \textsc{DistortionInspection} ($F_{4, 44} = 4.236$, $p < .01$). For follow-up analysis, we divide the trials into three groups based on  \textsc{DistortionAmount} and run one-way ANOVA on each group, investigating the influence of \textsc{DistortionInspection} on task completion time. For all groups, we find no significant main effect found (\textit{Low distortion}: $F_{2, 141} = 1.368$, $p = 0.258$; \textit{High MN}: $F_{2, 141} = 0.340$, $p = 0.712$; \textit{High FN}: $F_{2, 141} = 0.779$, $p = 0.461$).

\boldsubsubsection{Visual analysis of baseline techniques' task accuracy}
We visually explored additional factors affecting the interaction between \textsc{Techniques} and \textsc{DistortionAmount} (\autoref{fig:exp1_results}b). 
Although the difference is not statistically significant, we observe that the median task accuracy of $M$-Ball brushing (green) is higher than the one of Data-driven brushing (blue) for the projections with \textit{High FN}, while the opposite happens for the ones with \textit{High MN}.

\subsubsection{Discussions}

\label{sec:study1discuss}

We discuss the takeaways from our main findings (\autoref{sec:study1results}). 

\boldsubsubsection{\brush is more accurate than competitors when more distortions exist}
Findings A and B verify that \brush significantly outperforms baseline techniques, showing substantial accuracy regardless of distortions. 
These findings clearly imply the benefit of using \brush in MD data analysis. 
\brush will help analysts to reliably analyze the cluster structure of MD data even if MDPs have unreliable representations (\autoref{sec:intro}, \ref{sec:distortions}).

\boldsubsubsection{Baseline techniques can still be used in low-distortion MDP}
Our findings also indicate that baseline techniques can still be useful in some situations. 
Finding B-1 informs that baseline techniques are reliable if we can ensure that the projections have low distortions. In such a case, the analysis will benefit from the faster completion time of baseline techniques (Finding C). 

\boldsubsubsection{Local and global inspections are helpful when distortions are high}
The study does not reveal a significant influence on the existence of our inspection functionalities (\hspace{1pt}\stepone{} \steptwo{}\hspace{1pt}). 
However, in the interaction effect analysis on task completion time, we can observe that $F$-value of ANOVA decreases for \textit{High MN} and \textit{High FN} case, which means that the influence of global and local inspection increases when MDPs suffer from more distortions. Although not supported by statistical evidence, this observation aligns with the qualitative feedback from participants (\autoref{sec:qualinspection}) that our global and local inspection functionalities help users perform brushing tasks faster.

\boldsubsubsection{Trade-off between task accuracy and task completion time}
Although \brush shows the best task accuracy, it trades task completion time for such achievement (Finding C, D). 
Our qualitative interview reveals that this tradeoff mainly originates from the enhanced concentration or cautiousness of participants while using \brush (\autoref{sec:qualbenefit}). 

\subsection{User Study 2: Robustness Against the Non-Triviality of Multidimensional Cluster Shape}

\label{sec:detailstudy}

\begin{figure*}
    \centering
    \includegraphics[width=\linewidth]
    {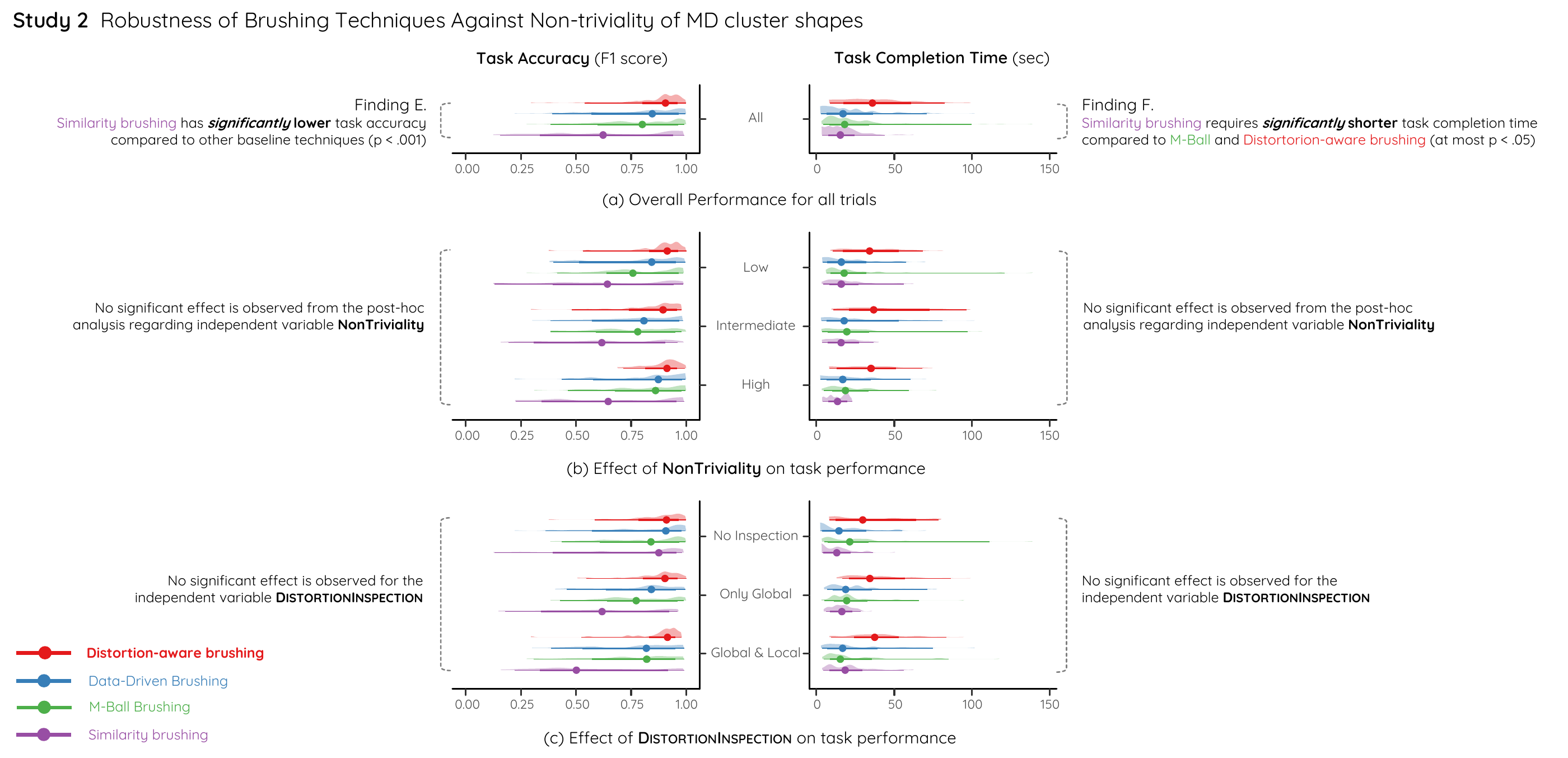}
        \vspace{-6mm}
    \caption{Study 2 results demonstrating the robustness of brushing techniques against the non-triviality of MD cluster shapes (\textsc{NonTriviality}). The study results reaffirm the superiority of \brush in terms of task accuracy. We find no significant differences between brushing techniques in terms of \textsc{NonTriviality}.}
    \label{fig:exp2_results}
\end{figure*}

\subsubsection{Objectives and Design}

We aim to investigate the robustness of \brush in maintaining user performances (task accuracy and completion time) against the non-triviality of cluster shapes.
We also want to check whether the general findings about the performance of \brush we find in Study 1 (\autoref{sec:comparisonstudy}) are maintained. 
To this end, the study has: 
\begin{itemize}
    \item Non-triviality of the MD shape of a designated cluster to be brushed (\textsc{NonTriviality})
    \begin{itemize}
        \item \textit{High}, \textit{Intermediate}, and \textit{Low},
    \end{itemize}
\end{itemize}
as an independent variable instead of \textsc{DistortionAmount}.
The amount of distortions is controlled for a confounding variable, along with the number of points and clusters.
The study also shares the independent variables \textsc{GlobalInspection} and \textsc{Techniques}, and all other experimental settings with Study 1. 
In summary, this study is identical to the previous one except that we swapped \textsc{DistortionAmount} for \textsc{NonTriviality}, and we work with different sets of participants.

\boldsubsubsection{Participants} We recruit 12 participants (E2P1--E2P12) from two local universities (ten males and two females, aged 21--30 [24.6 $\pm$ 2.5]) with the same recruiting criteria with the Study 1.

\subsubsection{Quantitative Results}

\label{sec:exp2quant}

The following are the findings from the study (\autoref{fig:exp2_results}).
We first check whether our findings from the previous study which are not dependent on \textsc{DistortionAmount} are maintained in this study (Finding A and C). We also discuss new findings from the study.

\boldsubsubsection{Analysis on overall task accuracy}
By executing three-way RM ANOVA examining the influence of three independent variables on the F1 score, we find a significant main effect on \textsc{Techniques} ($F_{3, 33} = 34.51$, $p < .001$). Bonferroni post-hoc analysis reveals that \brush significantly outperforms all other techniques ($p < .001$ for all) in terms of accuracy, again confirming Finding A.
Unlike Study 1, post-hoc analysis indicates that Data-driven brushing and $M$-Ball brushing show significantly higher accuracy compared to Similarity brushing ($p<.001$ for both). This leads to a new finding:

\finding{E}{Using similarity brushing leads to lower accuracy compared to not only \brush but also Data-driven brushing and $M$-Ball brushing.}

\noindent
We find a significant main effect for \textsc{NonTriviality} (($F_{2, 22} = 8.55$, $p < .01$), but post-hoc analysis fails to identify significant differences between brushing techniques.

\boldsubsubsection{Analysis on overall task completion time}
We run three-way RM ANOVA to investigate how three independent variables affect task completion time (\autoref{fig:exp2_results} right), which is followed by Bonferroni correction for post-hoc analysis.

RM ANOVA reveals a significant main effect on \textsc{Techniques} variable (($F_{3, 33} = 13.84$, $p < .001$)). The post-hoc analysis indicates that \brush requires a significantly longer task completion time compared to all other techniques ($p < .001$ for all). This result again confirms Finding C from the previous study. Moreover, the post-hoc analysis finds that the task completion time with $M$-Ball brushing is significantly longer than Similarity brushing ($p < .05$), hence the finding:

\finding{F}{Using similarity brushing leads to shorter task completion time compared to not only \brush but also $M$-Ball brushing.}

\noindent
We again find a significant main effect for \textsc{NonTriviality} ($F_{2, 22} = 5.22$, $p < .05$), but post-hoc analysis identifies no significant differences between brushing techniques.

\subsubsection{Discussions}

Study 2 again confirms the superiority of \brush regarding task accuracy and the tradeoff between task accuracy and completion time.
Moreover, it also reaffirms that there are no significant effects relevant to \textsc{DistortionInspection} (See \autoref{sec:study1discuss} for detail). The following is a takeaway from the new findings of Study 2 (\autoref{sec:exp2quant}).

\boldsubsubsection{Superiority and Inferiority between baseline techniques}
Finding E shows that $M$-Ball Brushing and Data-Driven Brushing have better task accuracy than Similarity brushing. Finding F then shows that $M$-Ball brushing requires longer task completion time compared to Similarity brushing.

Again, these results are consistent with the techniques' design. $M$-Ball brushing progressively and manually expands the brush while avoiding FN distortions, trading time for accuracy. In contrast, Similarity brushing selects all at once the true neighbors of all the points in the painter, which are not necessarily true neighbors of each other (FN), hence trading accuracy for time. Data-driven brushing also captures FN but does not automatically extend the brush to its true neighbors, resulting in fewer distortions than Similarity brushing. 
Our interview results (\autoref{sec:qual}) further verifies the finding that Data-driven brushing showed better task accuracy compared to Similarity brushing. 

%

\subsection{Post-Study Interview}

\label{sec:qual}

We discuss the post-study interview results and takeaways.

%



\subsubsection{Benefits in Terms of User Experience}

\label{sec:qualbenefit}

Our interview reveals \brush's positive effects on user confidence, cautiousness, and serendipity. 

\boldsubsubsection{\brush makes users more confident}
Participants report that \brush makes them more confident. 
 Specifically, 20 out of 24 participants (83\%) report that they are most confident when using \brush.
 
We find that such benefit mainly originates from the fact that \brush allows users to make final decisions about adding new points to the brush (\hspace{1pt}\stepfour{}\hspace{1pt}). In contrast, baseline techniques automatically formulate the MD region and only ``notify'' users of the points within the region being brushed points. 10 out of 24 participants (42\%) explicitly claim they are not able to manually fine-tune the brushed points in the baseline techniques, which leads them to lower confidence. 

We also find that participants feel more confident when using \brush as they can see an MD cluster as a 2D cluster (\hspace{1pt}\oone{}\hspace{1pt}). Eight out of 24 (33\%) participants mention that they are more confident about their interactions as the visual 2D clusters inform that they are doing well. 
In contrast, participants are not able to visually confirm the correctness of their interaction in baseline techniques, which makes them less confident.


\boldsubsubsection{\brush makes users more cautious}
Our interview reveals that the enhanced confidence again derived from the enhanced cautiousness of participants. E2P5 noted, `\textit{`As I knew that I was doing well, I wanted to perform my best in performing the given task''}. 
The result also aligns well with our quantitative findings. It is intuitive that people will require more time to complete tasks when they are more cautious (Finding C), but this will result in better accuracy (Findings A, B). 

\boldsubsubsection{Serendipity of \brush}
An unexpected benefit of \brush is that users enjoy using it. 11 out of 24 participants (45\%) note that playing with \brush is fun, and five among them (21\%) especially use the word ``game''. E1P6 noted, \textit{``I feel like playing a game that captures small monsters. I want to achieve a higher score.''}.



\subsubsection{Benefits of Global and Local Inspections}

\label{sec:qualinspection}

Although not statistically validated, our studies provide evidence that global and local inspection functionalities (\hspace{1pt} \stepone{} \steptwo{}\hspace{1pt}) provide a positive effect in reducing task completion time 
(\autoref{sec:study1discuss}). Aligned with this result, our interview also provides qualitative evidence of the positiveness of local inspection functionality. Seven out of 24 participants (29\%) reported that the local inspection functionality substantially helps them quickly determine which action to execute next. However, no participants explicitly indicate the benefit of global inspection functionality. We believe this is primarily because image snippets already support easy identification of the correct starting point for brushing. Examining the effect of global inspection without image snippets will be an interesting future avenue to explore.

\subsubsection{Limitations}

\label{sec:techlimit}

The interview also reveals the limitation that \brush requires an initial learning phase.
Participants report that it is not cognitively difficult to understand the workflow, but it still takes some time to become proficient with the technique. Sixteen out of 24 participants (67\%) mention that some practice is necessary to fully understand and utilize the functionalities of \brush. 
\revise{This is partly because participants are less familiar with painter-based brushing. Ten out of 24 participants (42\%) find the baseline Data-driven brushing more comfortable, as they are familiar with box-shaped brushes.}
\revise{We discuss strategies to make \brush more intuitive for first-time users in \autoref{sec:limitations}.}
\section{Use Case 1: \revise{Geospatial Cluster Analysis}}

\label{sec:scenario}

We demonstrate a use case validating \brush's effectiveness in \revise{conducting cluster analysis of geospatial data}.

\begin{figure*}[!ht]
  \centering
  \includegraphics[width=\linewidth]{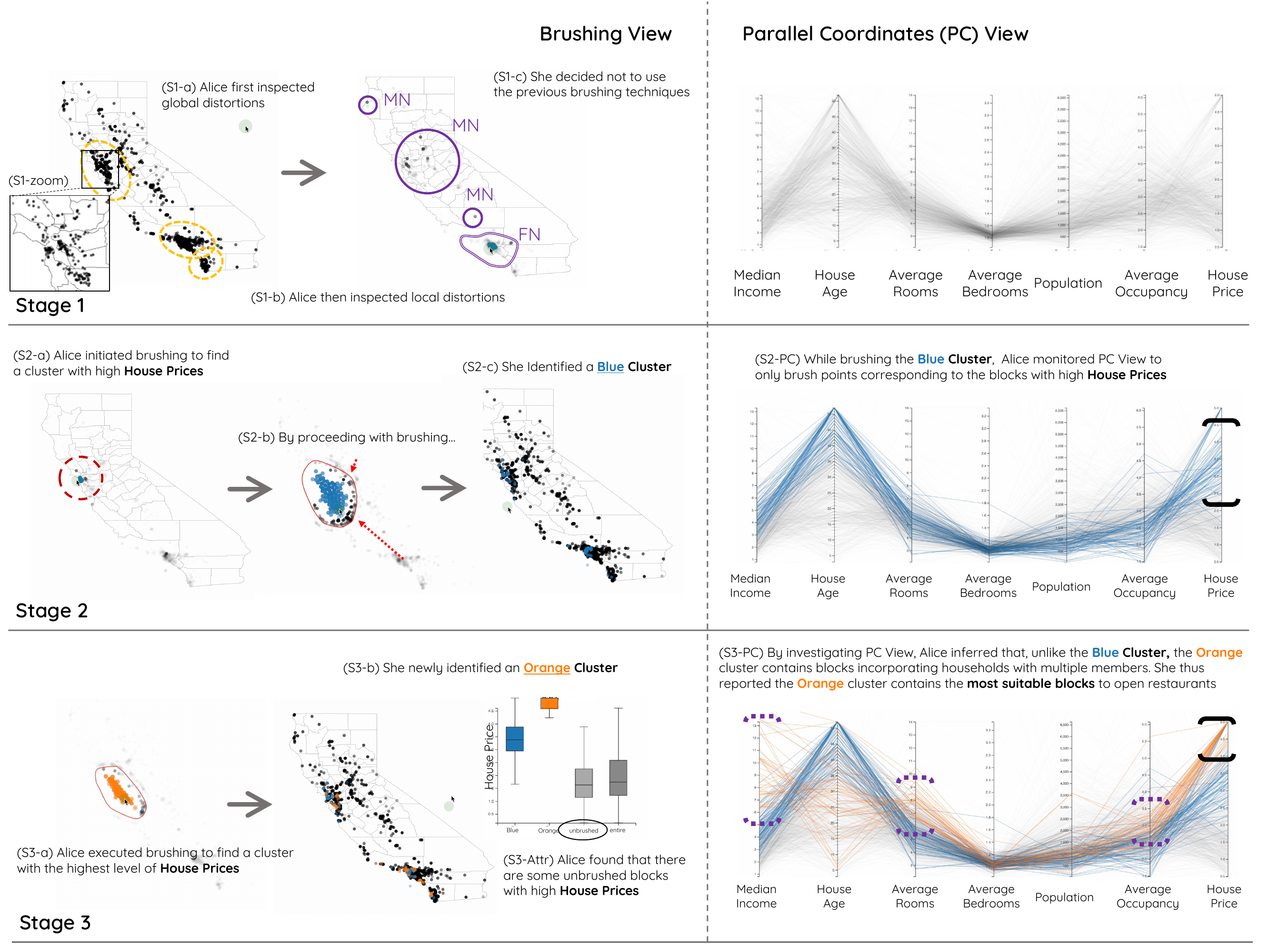}
  \vspace{-6mm}
  \caption{Use Case 1: Alice 
  explores the California Housing Dataset \cite{pace97statistics} (\autoref{sec:scenario}) with \brush to find a good candidate region for opening casual dining restaurants. 
 Alice uses a visual analytics system composed of a brushing view (left) that serves \brush and a Parallel Coordinates (PC) view (right) that shows the attribute values of brushed points. 
 Alice also checks detailed statistics of attribute values in brushed points by checking the Attribute (Attr) view.
 Using a visual analytics tool incorporating \brush, our persona, Alice, extracted clusters of blocks with similar characteristics and successfully conducted the given request.
  }
  \label{fig:scenario}
\end{figure*}

\subsection{Procedure}

\noindent
\textbf{Persona and Goals.}
We define a persona named Alice, a data analyst hired by a casual dining franchise company. The company wants to open new restaurants in California, but investigating the profitability of every block in the state exceeds its budget. Thus, the company asks Alice to find good candidate areas for detailed examination with three constraints: \textbf{(C1)} report blocks that are geographically similar to each other, which will substantially reduce the investigation cost. \textbf{(C2)} report blocks that are similar to each other for many attributes (i.e., well clustered in the MD space), also for reducing the investigation cost. 
\textbf{(C3)} focus on the blocks with higher average home prices, as people who have the financial means to pay for the restaurant’s food may live in such areas.

\noindent
\textbf{Dataset.} 
Alice uses California Housing dataset \cite{pace97statistics}, comprised of nine attributes (\textit{Longitude}, \textit{Latitude}, \textit{Median Income}, \textit{House Age}, \textit{Average Rooms}, \textit{Average Bedrooms}, \textit{Population}, \textit{Average Occupancy}, and \textit{House Price}). Each datum corresponds to an individual block, which is the smallest geographical unit used in the U.S. census. Each attribute value is a statistic that summarizes all households in a corresponding block. \reviset{Note that we randomly sample 5\% (1031 points) of the original dataset provided by scikit-learn \cite{pedregosa11jmlr} as too many data points may result in visual clutter in projections (\autoref{sec:scalabilitydisc}).}

\noindent
\textbf{System design.}
Alice uses a visual analytics system leveraging \brush consisting of three components: Brushing View, Parallel Coordinate (PC) View, and Attribute View. 

Brushing View provides \brush, where 2D projection is made by mapping longitude and latitude to $x$ and $y$ axes, and the map of California is displayed in the projection background. The MD space is set as the 7D space formed by all other attributes. The map is hidden when the points are relocated (Steps 3, 4), as the 2D positions of data points have no more direct relation to the map. When users return to the original projection using contextualization (\autoref{sec:contextualization}), they can see the map again. 

PC View helps users to directly monitor how the brushing proceeds and get hints about initiating or terminating brushing. PC dynamically reacts to user interaction in the Brushing View; the lines corresponding to seed points while inspecting local distortions (\hspace{1pt}\steptwo\hspace{1pt}), and the brushed points (\hspace{1pt}\stepfour\hspace{1pt}) are highlighted using the same color as the points. All other points are depicted in black with lower opacity.  


Finally, the Attribute (Attr) View displays the distribution of attribute values of (1) brushed clusters, (2) remaining points, and (3) all points using boxplots. Boxplots are dynamically updated, reflecting the brushing status in the Brushing View. 

\subsection{Scenario}

\subsubsection*{(Stage 1) Inspecting local distortions}

Alice first wants to examine whether she could report geographically similar points as clusters (C1). To do so, she checks whether visual proximity between 2D points matches the MD similarity. By inspecting the global distortion using density encoding (\hspace{1pt}\stepone\hspace{1pt}), she finds several 2D clusters in the projection (i.e., map) that have high MD density (yellow dotted ellipses in \autoref{fig:scenario} S1-a). 
Note that she zooms in to remove clutter and verifies there are points with high MD density in such places (\autoref{fig:scenario} S1-zoom).
Thus, she hypothesizes these regions may contain blocks that can also be considered clusters in the MD space. 

To validate her hypothesis, Alice hovers the painter around three regions to examine their local distortion. As a result, she finds that all regions suffer from both FN (neighboring map locations having low MD similarity, depicted by low opacity; purple doubled line in \autoref{fig:scenario} Stage 1) and MN (map locations far apart having similar attributes; purple solid line in \autoref{fig:scenario} Stage 1), rejecting the hypothesis (\autoref{fig:scenario} S1-b). This means that Alice \textit{cannot use conventional brushing or previous brushing techniques for MDP} (\autoref{sec:rel}) as they define the 2D brush as a compact, continuous 2D region vulnerable to FN. Therefore, Alice reports that it is difficult to satisfy C1 and decided to keep using \brush for the following analysis (\autoref{fig:scenario} S1-c).

\subsubsection*{(Stage 2) Extracting a cluster with high house prices}

Alice then aims to find points having high \textit{House Prices} (C3) and being well clustered (C2). 
She first moves the painter around the projection while monitoring seed points’ attribute values using PC view. As she expects that such ``wealthy regions'' may be located in urban areas, she especially examined regions around Los Angeles and San Francisco. As a result, she finds an area that contains seed points having \textit{House Prices} ranging from 250,000\$ to 400,000\$, which exceeds the median house price, near San Francisco downtown (red dashed circle in \autoref{fig:scenario} Stage 2; S2-a). She starts brushing from the seed points while monitoring the PC View (Steps 3, 4) (\autoref{fig:scenario} S2-b), and terminates brushing when the brush starts to contain the blocks that have significantly different patterns in PC View. As a result, she brushes the blocks with \textit{House Prices} lower than approximately 200,000\$ (black solid bracket in \autoref{fig:scenario} S2-PC). Following the corresponding brush color, she names the cluster as a \textit{Blue cluster} (\autoref{fig:scenario} S2-c). Note that this \textit{cannot be done by filtering high-price blocks in PC View} as she does not know the threshold of \textit{House Prices} that can accurately discriminate blocks having different patterns in PC View.


\subsubsection*{(Stage 3) Extracting a cluster with the highest house prices}

Through the PC View and the Attribute View, Alice discovers that the portion of blocks with the highest \textit{House Prices} is not incorporated in the Blue cluster.
Therefore, she decides to find data points with similar attribute values (C2) that may comprise these highly-priced blocks (C3). By skimming through the urban area again, she finds seed points with the highest level of \textit{House Prices} and starts brushing again (\autoref{fig:scenario} S3-a). To clearly discriminate the currently brushed cluster from the Blue cluster, she stops brushing when the newly brushed cluster starts to incorporate the blocks with prices lower than 400,000\$ (black solid bracket in \autoref{fig:scenario} S3-PC). She names the cluster the \textit{Orange cluster}, following the same naming convention (\autoref{fig:scenario} S3-b).

Alice finds that the orange cluster not only exceeds the blue cluster in terms of \textit{House Prices} but also in \textit{Median Income}, \textit{Average Room}, and \textit{Average Occupancy} (purple dotted brackets in \autoref{fig:scenario} S3-PC). Higher \textit{Average Occupancy} denotes that the blocks in the orange cluster may contain households consisting of multiple people. Higher \textit{Median Income} and \textit{Average Room} also support this inference, as they will grow proportionally to the number of members in each household. Alice thus concludes that the \textit{Orange cluster} contains blocks that are more appropriate for opening casual dining restaurants, as such restaurants may not target people who eat alone. 

Alice thus reports the blocks within the orange cluster as having top priority for investigation. It is worth noting that blue and orange clusters have no clear segmentation regarding individual attribute values (\autoref{fig:scenario} S3-PC). This means that the \textit{neither brushing axes of the PC nor previous brushing techniques for MDP cannot precisely reveal these clusters} (see Appendix G for further confirmation). Moreover, points within two clusters are all scattered throughout the projection (\autoref{fig:scenario} Stage 4). This reaffirms that \textit{both naive 2D brushing and previous brushing techniques for MDP cannot stand out to detect these clusters}, although these clusters have clear differences overall (S3-PC)---yet a clear benefit of \brush in this scenario.

\begin{figure}
    \centering
    \includegraphics[width=\linewidth]{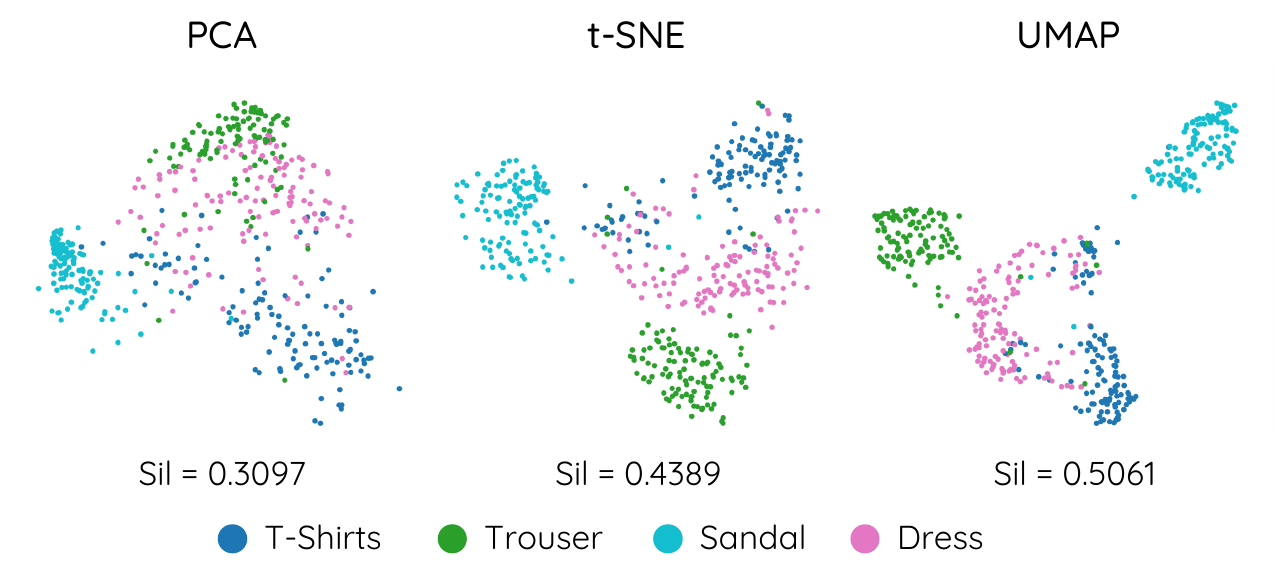}
    \vspace{-7mm}
    \caption{\revise{PCA, $t$-SNE, and UMAP projections of the subset of Fashion-MNIST dataset used in our second use case (\autoref{sec:usecaselabeling}). 
    Silhouette (Sil) scores indicate the separability of classes in the projections.
    Data points are assumed initially unlabeled, appearing as monochrome scatterplots.}}
    \label{fig:usecase2_projection}
\end{figure}

\section{\revise{Use Case 2: Visual-Interactive Labeling}}

\label{sec:usecaselabeling}

\revise{
\brush can effectively identify noisy clusters that are semantically distinct but not well-separated in the data space (\hspace{1pt}\stepfour{}\hspace{1pt}).
This enables users to make the final decision on whether to brush certain points.
We present a use case leveraging this utility in supporting the visual-interactive labeling \cite{bernard18tvcg, meng24tvcg} for data with noisy MD clusters and distorted projections. 
}

\subsection{Procedure}

\noindent 
\textbf{\revise{Persona and goals.}}
\revise{
Bob, a data engineer, wants to train a classification model on an image dataset. He aims to perform visual-interactive labeling to identify ground-truth classes.}

\begin{figure*}
    \centering
    \includegraphics[width=\textwidth]{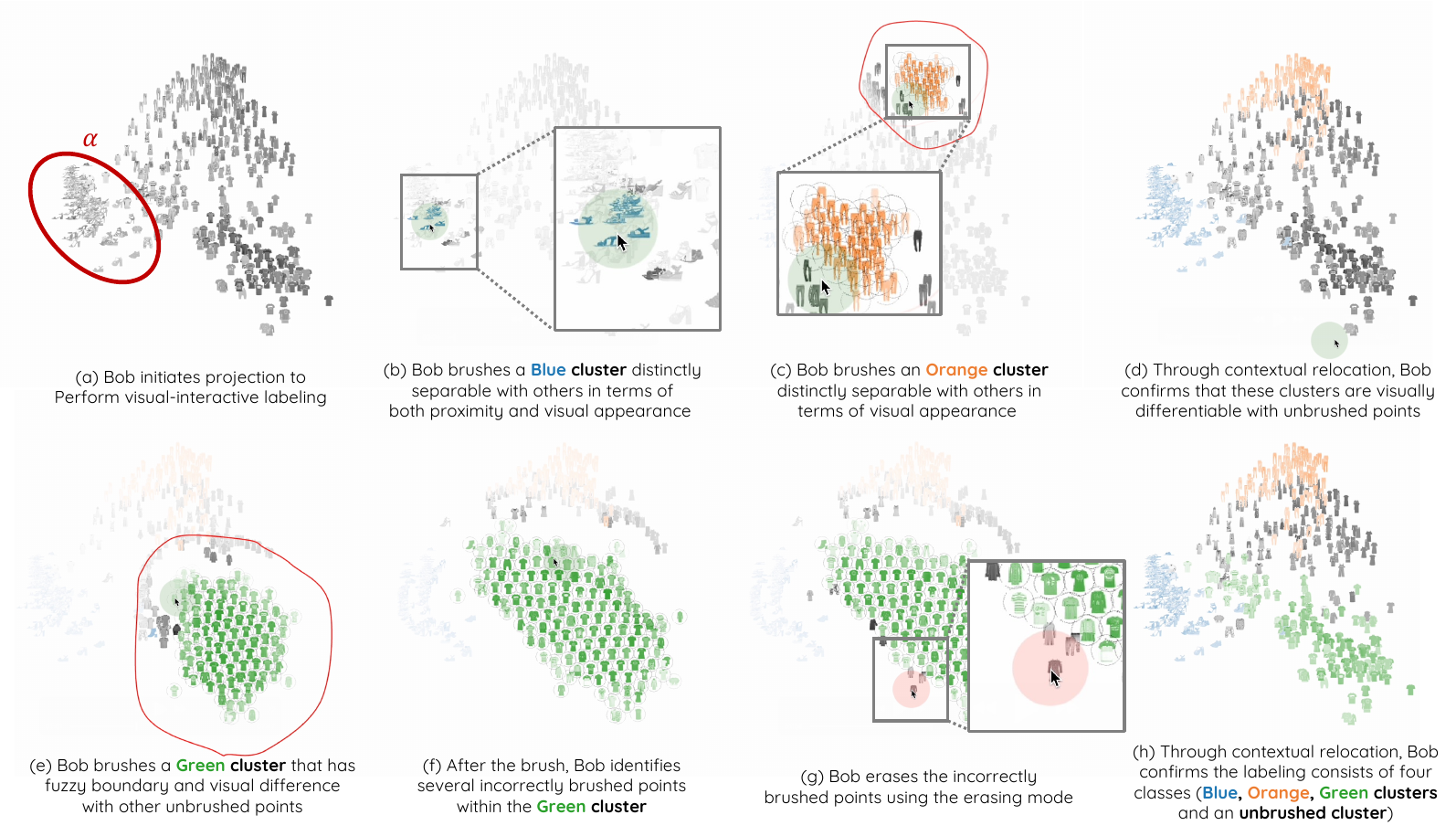}
    \vspace{-7mm}
    \caption{\revise{Use Case 2: Bob uses \brush for visual-interactive labeling \cite{bernard18tvcg, meng24tvcg} (\autoref{sec:usecaselabeling}). Bob aims to identify and brush semantically distinct clusters in the dataset to form ground-truth label data for training an automatic classifier. \brush is effective to identify noisy clusters (h) (semantically separated in snippet images but not in the data space) from strongly distorted projections (a).}}
    \label{fig:usecase2_scenario}
\end{figure*}

\noindent
\textbf{\revise{Dataset.}}
We simulate a dataset with noisy clusters, i.e., clusters that overlap in the data space, making them difficult to detect using automatic clustering algorithms or nonlinear dimensionality reduction techniques. We extract four classes in the Fashion-MNIST dataset \cite{xiao2017arxiv}: \textit{T-Shirts, Trouser, Sandal,} and \textit{Dress} (\autoref{fig:usecase2_projection}). Among them, \textit{T-Shirts} and \textit{Dress} have been verified to have low MD pairwise separability \cite{jeon24tvcg}. We assume that the dataset lacks predefined class labels, requiring Bob to identify these four classes.
As shown in \autoref{fig:usecase2_projection}, these classes are not well-separated in projection spaces. This implies that previous brushing techniques may poorly discriminate these classes (\autoref{sec:userstudy}).
\reviset{Note that we use the entire 784 dimensions of the original dataset to represent the MD space.}
\reviset{To avoid visual clutter (\autoref{sec:scalabilitydisc}), we use 2\% (120 points for each cluster) of the original Fashion-MNIST dataset.}

\noindent
\textbf{\revise{System design.}}
\revise{Bob uses a labeling system equipped with \brush. 
To simulate a scenario where projections distort the representation of MD clusters, we use PCA, which shows the worst performance in discriminating classes in terms of Silhouette \cite{rousseuw87silhouette} scores (\autoref{fig:usecase2_projection}).
Each data point is visualized as an image snippet; Bob can thus investigate the semantic difference between data points based on their visual appearance.
}

\subsection{\revise{Scenario}}

\subsubsection*{\revise{(Stage 1) Brushing a Cluster Separated by Visual Similarity and Proximity}}

\revise{
Upon initiating the system (\autoref{fig:usecase2_scenario}a), Bob identifies a cluster that is visually and spatially distinct from other data points (\autoref{fig:usecase2_scenario}$\alpha$; corresponds to the \textit{Sandal} class). Bob uses \brush to label this cluster. As the cluster is clearly separated from other points in both the MD space and the projection, Bob brushes the cluster with no difficulty  (\autoref{fig:usecase2_scenario}b), naming it the \textit{Blue Cluster}.
}

\subsubsection*{\revise{(Stage 2) Brushing a Cluster Separated By Visual Similarity}}
\revise{
Bob identifies that the remaining points consist of two types of fashion items: \textit{tops} (corresponding to \textit{T-Shirts} and \textit{Dress} classes) and \textit{bottoms} (corresponding to \textit{Trouser}). He notices that these two groups are not well-separated in the projection but have semantically distinct visual appearances. To differentiate them, Bob brushes the \textit{bottoms} (\autoref{fig:usecase2_scenario}c).  Although \textit{tops} and \textit{bottoms} overlap in the projection, \brush dynamically relocates the selected points, enabling Bob to discriminate between them with ease. 
He names this cluster the \textit{Orange Cluster}. By contextualizing the brushing results within the original projection (\autoref{sec:contextualization}), Bob confirms that the \textit{Blue Cluster} and \textit{Orange Cluster} represent distinct labels (\autoref{fig:usecase2_scenario}d).
}

\subsubsection*{\revise{(Stage 3) Separating Clusters with Fuzzy Boundaries}}
\revise{Bob inspects the remaining points that correspond to \textit{tops} and identifies that they consist of two semantically different clusters (\textit{T-shirts} and \textit{Dress} classes). 
However, these two clusters are semantically not clearly separated and partially overlapped in the projection. Despite this, Bob brushes one of them, naming it the \textit{Green Cluster} (\autoref{fig:usecase2_scenario}e). As \brush relocates points, it helps Bob consolidate semantically similar points scattered throughout the projection. However, after completing the brushing, he observes that the \textit{Green Cluster} still contains points that do not appear to belong to the correct label (\autoref{fig:usecase2_scenario}f). To refine the selection, he activates the erasing mode (\autoref{sec:erasing}) and removes these points (\autoref{fig:usecase2_scenario}g). Since uncertain points are pushed to the cluster boundary \hspace{1pt}\stepfour\hspace{1pt}), Bob erases them with ease.
}

\revise{
Finally, by once again contextualizing the brushing results within the original projection (\autoref{fig:usecase2_scenario}h), Bob confirms that the labeling now consist of four clusters (\textit{Blue, Orange, Green} and the remaining \textit{unbrushed} clusters).
}

\subsection{\revise{Comparison with Existing Systems}}

\revise{Previous visual-interactive labeling systems \cite{bernard18tvcg, meng24tvcg} provide an initial projection and prompt labelers to label a subset of the data. These labeled points are then incorporated into an underlying classification model, which supports the labeling process by (1) predicting labels for unlabeled points \cite{bernard18tvcg}, or (2) updating projections to better highlight the separation of labeled points \cite{meng24tvcg}. This iterative process continues until labeling is complete.
}

\revise{
Our \brush approach offers two main advantages over the previous systems. First, in the earlier systems, the initial labeling can be erroneous because of distortions in the initial MDP, and these errors can propagate through subsequent iterations as the labels are used to train the classification model. Our approach avoids this problem through point relocation. Second, our approach does not require an underlying classification model at all, which simplifies the system by reducing hyperparameters and computational overhead. This also helps users to preserve their mental map throughout the labeling process, as they do not need to wait for model training at each iteration.
However, when MDPs have less distortion and clusters are not noisy (i.e., aligned well with potential classes), \brush may provide little advantages over previous systems. 
}
\revise{Further quantitative comparisons are needed to better evaluate the pros and cons of both approaches.}





\section{Discussions}

\subsection{Comparison to Automatic Clustering Techniques}

\brush and automatic clustering techniques (e.g., $K$-Means) both aim to discover meaningful cluster patterns in MD data. 
However, by engaging users in the cluster extraction process, \brush makes cluster analysis more insightful. 
First, by allowing users to ``see'' image snippets (\autoref{fig:teaser}) or auxiliary visualizations (\autoref{sec:scenario}), \brush can account for visual similarity between data points that may often overlooked by clustering techniques that rely on conventional similarity or distance metrics (e.g., Euclidean, $k$NN). 
\revise{Our second use case (\autoref{sec:usecaselabeling}) demonstrates the benefits of this approach.}
Moreover, as seen in our first use case (\autoref{sec:scenario}), \brush supports users to incorporate their external knowledge into the clustering procedure, specify characteristics of clusters that they want to discover, and set the boundary of clusters themselves. These functionalities enrich data analysis and are not possible in conventional clustering techniques.
Such benefits emphasize both the effectiveness of \brush and the importance of the interactive aspect in cluster analysis.

\subsection{Additional Usage Scenarios}

This research contributes \brush to make \textit{cluster analysis of MD data more reliable}. Our studies and use cases align with this purpose. 

However, the core concept of \brush can be applied to exploring and clustering diverse data entities where discrepancies exist between their 2D spatial positions and semantic meanings. 
For example, \brush can be used to browse computer files within messy directories (e.g., \texttt{Downloads} or \texttt{Desktop}) by interactively clustering the files based on content similarity.
In this case, users would not only rely on the snippet image of each file (i.e., icons) but also benefit from invisible semantic distances between file contents, with distortions computed accordingly to guide the \brush.
\revise{
Another potential use case of \brush is to compare the hyperparameter settings of machine learning models, where \brush could help users compare different hyperparameter settings, identify a group 
of optimal configurations, and gain insight into how their variations affect model performance.}


\subsection{Visual and Computational Scalability}

\label{sec:scalabilitydisc}

The computational \reviset{time} complexity of a single iteration of lens update and point relocation is $O(m\log m + \kappa nm)$, where $n$ denote the total number of points, $m$ is the size of the convex hull, and $\kappa$ denotes the number of nearest neighbors considered to compute closeness (we detail the complexity analysis in Appendix C). As $n \gg m$ and $n \gg \kappa$, the running time of the technique is mainly bounded by $n$ \reviset{and has no relation with the dimensionality}, making it highly scalable. 
\reviset{We empirically find that \brush ensures responsive interaction to the datasets that have approximately 60,000 data points (Appendix J). Note that the preprocessing of our technique requires up to $O(n^2k^2)$ but needs to be executed only once and does not impact the interactivity.}

However, the technique may still \reviset{struggle to deal with large datasets for two reasons. First, the technique may suffer from} visual complexity.
When applied to a large dataset, visual clutter can hide image snippets and opacity encoding. 
Enhancing \brush's encoding to mitigate such visual complexity will be an interesting future work. For example, we may adopt density plots \cite{trautner2020sunspotPlots} to resolve the clutter. 
\reviset{Second, the technique requires a substantial amount of memory. We store the similarity matrix in dense format, requiring $O(n^2)$ memory, which causes an out-of-memory error in the heap space with large datasets (Appendix J). Reducing memory usage, for example, by utilizing sparse matrix formats, will be crucial for enhancing the scalability of \brush when processing large datasets.}


\subsection{\revise{Reproducibility of Cluster Analysis}}

\revise{
Reproducibility is a fundamental aspect of visual analytics. Since \brush allows analyses to start from different points, potentially leading to varying outcomes (\autoref{sec:contextualization}), it might seem to harm reproducibility. However, our technique mitigates this concern by guiding users in identifying MD clusters that align with 2D clusters, e.g., through opacity-coding of MD density (\hspace{1pt}\stepone\hspace{1pt}), which encourages users to initiate clustering in similar regions and reduces confirmation bias. Moreover, \brush is intentionally designed with minimal hyperparameters (\ofour), reducing variability from extensive hyperparameter tuning, which is a common pitfall in automated clustering algorithms. Finally, our user study demonstrates that \brush enables participants to identify clusters that align closely with ground-truth MD clusters, outperforming conventional brushing approaches. Collectively, these factors verify that \brush contributes to enhancing the reproducibility of MD cluster analysis.
}

\subsection{Limitations}

\label{sec:limitations}

\boldsubsubsection{Handling soft clusters}
In this research, we assume ``hard clusters'': every point can be assigned to at most one cluster. In reality, soft clusters are often observed; a point can belong to multiple clusters with a certain degree. 
Enhancing the technique to handle soft clusters will be an interesting future research.

\boldsubsubsection{\revise{Preserving mental map}}
While point relocation helps robustly brush MD clusters regardless of MDP distortions (\hspace{1pt}\otwo\hspace{1pt}), it comes at the cost of global context provided by the original MDP. \revise{As a result, users may struggle to maintain their mental map,} making it harder to perceive the connection between the original projection and the brushed regions. This can potentially hinder tasks that rely on global structure investigation, such as comparing the density of clusters \cite{xia22tvcg}. One possible solution is to juxtapose the interactive projection with the original MDP, trading visual space for better contextual awareness. 
\revise{
Our contextualization functionality (\autoref{sec:contextualization} further supports users in recalling the global structure by helping them maintain their mental map throughout the brushing process  
}

Still, by compromising on global reliability, users can gain substantially high reliability in extracting and examining local clusters, as demonstrated in our user study (\autoref{sec:userstudy}). 
This advantage enables \brush to reliably support the typical data analysis flow of the visual seeking mantra \cite{shneiderman03civ}: \textit{Overview first, zoom and filter, details on demand}. After users inspect the \textit{overview} of data distribution through the original MDP, they can accurately \textit{zoom into} or \textit{filter} local clusters using the technique, then reliably analyze clusters in \textit{detail} through image snippets or auxiliary visualizations (\autoref{sec:scenario}, \ref{sec:usecaselabeling}).

The strategy to enhance local reliability by sacrificing global reliability is similar to that of Focus-and-Context (FC) (e.g., Fisheye~\cite{gansner05tvcg}). However, while FC approaches only provide transient enhancement of local reliability, \brush continuously maintains the brushed local MD cluster, which enables more detailed and tangible follow-up analysis (\autoref{sec:scenario}).

\boldsubsubsection{\revise{Learning curve}}
\revise{
Our user study reveals that \brush takes time for users to become comfortable with (\autoref{sec:techlimit}). 
Post-hoc interviews suggest this is partly due to users' limited familiarity with painter-based brushing systems.
We plan to redesign \brush to leverage the box-shaped brush regions, which are more familiar to most users. 
Another possible approach is to slow down point relocation, allowing users to control when to perform lens updates and point relocation. While this may slightly increase task completion time, it could help users adapt to \brush more easily. 
}

\boldsubsubsection{Limitations in user studies}
We discuss the limitations of our studies. 
First, data points are visualized using image snippets in our experiment, but data points may lack intuitive visual representations, \textit{e.g.} in tabular data or text documents.
We plan to improve the generalizability of our study by testing the situation in which auxiliary visualizations (e.g., \autoref{sec:scenario}) are used to guide the brushing. 
We also want to clarify the need to conduct experiments with more datasets. For instance, \textsc{NonTriviality} may not have a statistically significant effect on the studied brushing techniques because the variation in the non-triviality of cluster shapes between digits in the MNIST dataset was not large enough.




\section{Conclusion}

Although brushing in MDP has long been considered an important research topic, previous brushing techniques for MDPs have struggled to overcome distortions. 
To tackle this problem, we proposed \brush, a brushing technique that relocates points to resolve distortions.  
Our user studies demonstrated the usefulness and usability of \brush in exploring and discovering MD clusters, its robustness to distortions, and its greater accuracy compared to state-of-the-art brushing techniques. 
In summary, our work advances the research community one step closer to making more reliable MD data analysis.

\section*{Acknowledgments}
\noindent This work was supported by the National Research Foundation of
Korea (NRF) grant funded by the Korea government (MSIT) (No. 2023R1A2C200520911) and the SNU-Global Excellence Research Center establishment project.
This work was also supported by the
Institute of Information \& communications Technology Planning
\& Evaluation (IITP) grant funded by the Korea government (MSIT)
[NO.2021-0-01343, Artificial Intelligence Graduate School Program
(Seoul National University)]. 
The ICT at
Seoul National University provided research facilities for this study.
Hyeon Jeon is in part supported by Google Ph.D. Fellowship. 
Ghulam Jilani Quadri was supported in part by NSF CNS \#2127309 to the Computing Research Association for the CIFellows Project.

\bibliographystyle{IEEEtran}
\bibliography{ref}

\vspace{-14mm}
\begin{IEEEbiography}[{\includegraphics [width=1in,height=1.25in,clip, keepaspectratio]{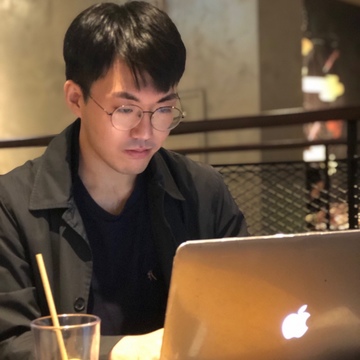}}] {Hyeon Jeon} 
is a Ph.D. Student at the Department of Computer Science and Engineering, Seoul National University. His research interests span the field of Visual Analytics and Machine Learning. Before starting his Ph.D. program, he received a B.S. degree in Computer Science and Engineering from POSTECH. 
\end{IEEEbiography}
\vspace{-11mm}
\begin{IEEEbiography}[{\includegraphics [width=1in,height=1.25in,clip, keepaspectratio]{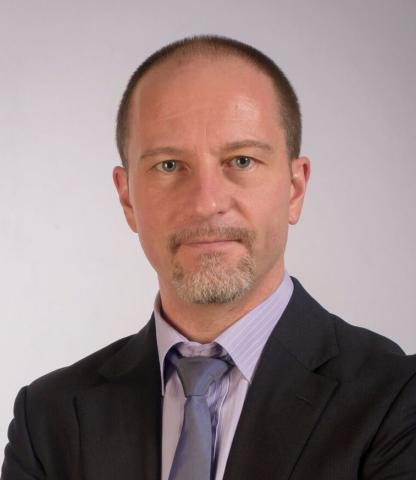}}] {Micha\"el Aupetit}  is a Senior Scientist
at the Qatar Computing Research Institute. He graduated with a Computer Science Engineering degree and an MSc in Robotics and Microelectronics from the University of Montpellier. He earned a PhD degree in Industrial
Engineering from Institut National Polytechnique
de Grenoble (INPG) in 2001, and an HDR in CS from the University of Paris-Saclay in 2011.
\end{IEEEbiography}
\vspace{-11mm}
\begin{IEEEbiography}[{\includegraphics [width=1in,height=1.25in,clip, keepaspectratio]{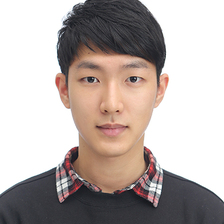}}] {Soohyun Lee} 
is a Ph.D. Student at the Department of Computer Science and Engineering, Seoul National University. Before starting his Ph.D. program, he received a B.S. degree in Computer Science and Engineering from the Korea University, Seoul, Korea. 
\end{IEEEbiography}
\vspace{-11mm}
\begin{IEEEbiography}[{\includegraphics [width=1in,height=1.25in,clip, keepaspectratio]{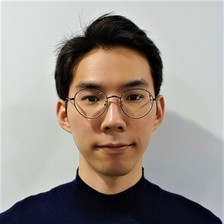}}] {Kwon Ko} 
is a Ph.D. Student at Stanford University. Prior, he received a B.S. degree in Mathematics from Hanyang University and received an M.S. degree in Computer Science and Engineering from Seoul National University.
\end{IEEEbiography}
\vspace{-11mm}
\begin{IEEEbiography}[{\includegraphics [width=1in,height=1.25in,clip, keepaspectratio]{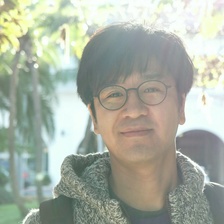}}] {Youngtaek Kim} 
is a staff engineer at Samsung Research. Prior to joining Samsung, he received a Ph.D. degree in Computer Science and Engineering from Seoul National University. 
\end{IEEEbiography}
\vspace{-11mm}
\begin{IEEEbiography}[{\includegraphics [width=1in,height=1.25in,clip, keepaspectratio]{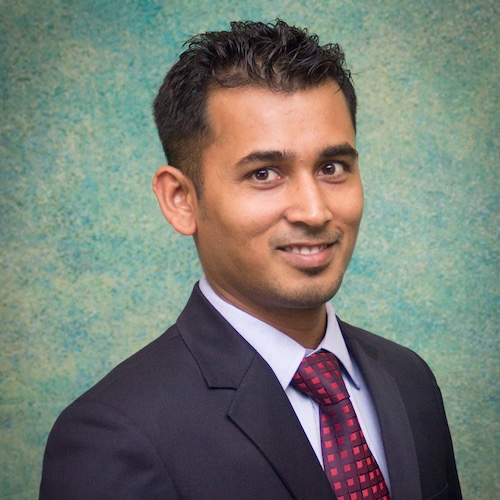}}] {Ghulam Jilani Quadri} received the Ph.D. degree from the University of South Florida. He is currently an assistant professor at the University of Oklahoma. 
Before joining the University of Oklahoma, he was a CIFellow postdoc with the Department of Computer Science, University of North Carolina
Chapel Hill.
\end{IEEEbiography}
\vspace{-11mm}
\begin{IEEEbiography}[{\includegraphics [width=1in,height=1.25in,clip, keepaspectratio]{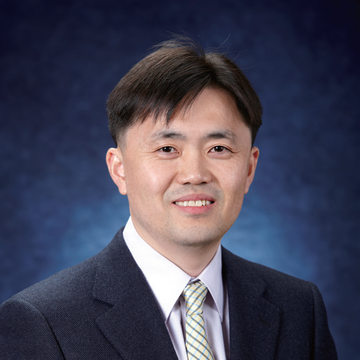}}] {Jinwook Seo} 
is a professor in the Department of Computer Science and Engineering, Seoul National University, where he is also the Director of the Human-Computer Interaction Laboratory. His research interests include Human-Computer Interaction, Information Visualization, and Biomedical Informatics. He received his PhD in Computer Science from the University of Maryland at College Park in 2005.
\end{IEEEbiography}

\end{document}